\documentclass[twocolumn]{aastex63}
\usepackage{lineno}
\pdfoutput=1 
\usepackage{amsmath,amstext,amssymb}
\usepackage[T1]{fontenc}
\usepackage[figure,figure*]{hypcap}
\usepackage[caption=false]{subfig}
\graphicspath{{./}{figures/}}

\received{?}
\revised{?}
\accepted{?}
\submitjournal{APJ}

\shorttitle{Higher-Order MMRs}
\shortauthors{Keller et al.}

\begin{document}

\title{Higher-Order Mean-Motion Resonances Can Form in Type-I Disk Migration}

\correspondingauthor{Finnegan Keller}
\email{finnegan\_keller@asu.edu}
\author[0009-0000-7547-7776]{Finnegan M. Keller}
\affiliation{Department of Physics, Brown University, Providence, RI 02912, USA}
\affiliation{School of Earth and Space Exploration, Arizona State University, Tempe, AZ 85287, USA}
\affiliation{Institute for Astronomy, University of Hawai`i, 2680 Woodlawn Drive, Honolulu, HI 96822, USA}

\author[0000-0002-8958-0683]{Fei Dai}
\affiliation{Institute for Astronomy, University of Hawai`i, 2680 Woodlawn Drive, Honolulu, HI 96822, USA}
\affiliation{Division of Geological and Planetary Sciences,
1200 E California Blvd, Pasadena, CA, 91125, USA}
\affiliation{Department of Astronomy, California Institute of Technology, Pasadena, CA 91125, USA}

\author[0000-0002-9408-2857]{Wenrui Xu}
\affiliation{Center for Computational Astrophysics, Flatiron Institute, New York, USA}


\begin{abstract} 
\noindent Type-I disk migration can form a chain of planets engaged in first-order mean-motion resonances (MMRs) parked at the disk inner edge. However, while second- or even third-order resonances were deemed unlikely due to their weaker strength, they have been observed in some planetary systems (e.g. TOI-178 bc: 5:3, TOI-1136 ef: 7:5, TRAPPIST-1 bcd: 8:5-5:3). We performed $>6,000$ Type-I simulations of multi-planet systems that mimic the observed {\it Kepler} sample in terms of stellar mass, planet size, multiplicity, and intra-system uniformity over a parameter space encompassing transitional and truncated disks. We found that Type-I migration coupled with a disk inner edge can indeed produce second- and third-order resonances (in a state of libration) in $\sim 10\%$ and 2\% of resonant-chain systems, respectively. Moreover, the relative occurrence of first- and second-order MMRs in our simulations is consistent with observations (e.g. 3:2 is more common than 2:1; while second-order 5:3 is more common than 7:5). The formation of higher-order MMRs favors slower disk migration and a smaller outer planet mass. Higher-order resonances do not have to form with the help of a Laplace-like three-body resonance as was proposed for TRAPPIST-1. Instead, the formation of higher-order resonance is assisted by breaking a pre-existing first-order resonance, which generates small but non-zero initial eccentricities ($e\approx10^{-3}$ to 10$^{-2}$). We predict that 1) librating higher-order resonances have higher equilibrium $e$ ($\sim 0.1$); 2) are more likely found as an isolated pair in an otherwise first-order chain; 3) more likely emerge in the inner pairs of a chain. 
\end{abstract}


\section{Introduction} 
\label{sec: 1}

There is mounting evidence \citep[e.g.][]{MillsNature,Izidoro,Leleu2021,dai2023toi,Luque110067, dai2024prevalence, hamer2024kepler} that Kepler-like planets \citep[$\sim 0.1$AU, $<4R_\oplus$, $N_p\geq3$, ][]{Winn_Fabrycky,Zhu_Dong} could have formed initially in chains of mean-motion resonance (MMR) through Type-I (non-gap-opening) convergent disk migration \citep{Goldreich1979, Ward1997, Lin1986, Kley_2012, Izidoro, ogihara2018formation, Wong2024}. After the gas disk dissipates, the stability of a planetary system is no longer protected by eccentricity and inclination damping induced by planet-disk interactions \citep{Papaloizou2000}. Over time, orbital instability or other dynamical effects disrupt the initial resonances, leading to the predominantly non-resonant orbital architectures observed among mature Kepler-like planets \citep{Pichierri2020, Goldberg2022, Izidoro,Goldberg_stability,Li2024, Matsumoto2012, rath2022criterion}. 

The order of a $p$:$q$ MMR is defined by the difference of two integers ($p$ and $q$) involved ($|p-q|$, e.g. 3:2 is first-order, 5:3 is second-order, 8:5 is third-order). While the distinction may appear trivial at first, a careful treatment of the resonant Hamiltonian \citep{Murray, Hadden2019} shows that the strength of the MMR scales as orbital eccentricity raised to the power of the order ($\propto e^{|p-q|}$). This is because the perturbation near conjunctions in higher-order resonance partially cancel out \citep{Tamayo2024}. For most Kepler-like planetary systems, the orbital eccentricity is low \citep[$\lesssim$ 0.05 e.g.][]{Hadden2017,Xie}, so higher-order MMR could be more than an order of magnitude weaker than first-order MMR. Acting as a weak link in a chain of mean-motion resonances, higher-order MMR may play a role in the disruption of initially resonant Kepler-like systems \citep{dai2023toi}. 

Unless characterizing a specific system with an apparent higher-order resonance(s) \citep{MacDonald, Mills2016, Siegel2021, MacDonald, quinn2023confirming, lammers2024six, dai2023toi, Tamayo2017, coleman2019pebbles, childs2025observational}, previous disk migration simulations usually overlooked higher-order MMR as capturing planets into weaker higher-order resonances was assumed to be much more difficult than first-order MMR. Pioneering works by \cite{xiang2015evolutionary,Xu0217} showed that capturing a pair of planets into second-order resonances requires more specific circumstances but is still possible. Higher-order MMR formation is also favored by convergent migration, an even slower migration rate, near-unity planet-planet mass ratios, and low (but not zero, see Section \ref{sec: 4.7}) pre-resonance eccentricities \citep{Xu0217, Batygin2015}. A number of multi-planet planetary systems contain planet pairs near higher-order resonances, including Kepler-$29$ bc: 9:7 \citep{Fabrycky2012,Migaszewski_Kepler29}, TOI-178 bc: 5:3 \citep{Leleu2021}, TOI-1136 ef: 7:5 \citep{dai2023toi}, Kepler-138 cde: 5:3-5:3 \citep{Jontof-Hutter138}, and TRAPPIST-1 bcd: 8:5-5:3 \citep{Gillon,Huang_Ormel}. Higher-order resonances are known for some Solar System objects \citep{Murray}, such as those between Neptune and many Kuiper Belt Objects \citep[e.g.][]{volk2025machine, smirnov2025highly, Chiang2003} although there is a striking deficit of them in the asteroid belt \citep{demeo2013taxonomic}. 

These observations prompted us to investigate the formation of higher-order MMR in Kepler-like systems through Type-I disk migration. While previous works \citep{xiang2015evolutionary,Xu0217} studied second-order MMR capture of isolated planet pairs, we investigated both second-order and third-order resonances in systems with more than two planets. We also incorporated a disk inner edge in our migration simulations based on disk truncation at the magnetospheric boundary \citep{Masset_2006,Izidoro,Wong2024}, which is crucial for stopping migration before the planets fall onto the star and for converting divergent encounters into convergent ones. Moreover, our simulations included protoplanetary disks of surface densities as low as 10 g~cm$^{-2}$ at 1 AU \citep[two orders of magnitude smaller than the Minimum Mass Solar Nebula, MMSN,][]{Hayashi} to mimic transitional disks or truncated disks, which may be relevant to small planet formation \citep{choksi2020sub, Lee2016transitional}.

 In Section \ref{sec: 2}, we describe our numerical disk migration model. In Section \ref{sec: 3}, we present two case studies of how higher-order MMR are produced during Type-I migration. We discuss population-level results in Section \ref{sec: 4}. Finally, Section \ref{sec: 5} contains a summary of this paper.

\section{Methods}
\label{sec: 2}

\subsection{Disk Migration Setup} 
\label{sec: 2.1}

\begin{figure*}[ht] 
\begin{center} 
\includegraphics[width = 2.0 \columnwidth]{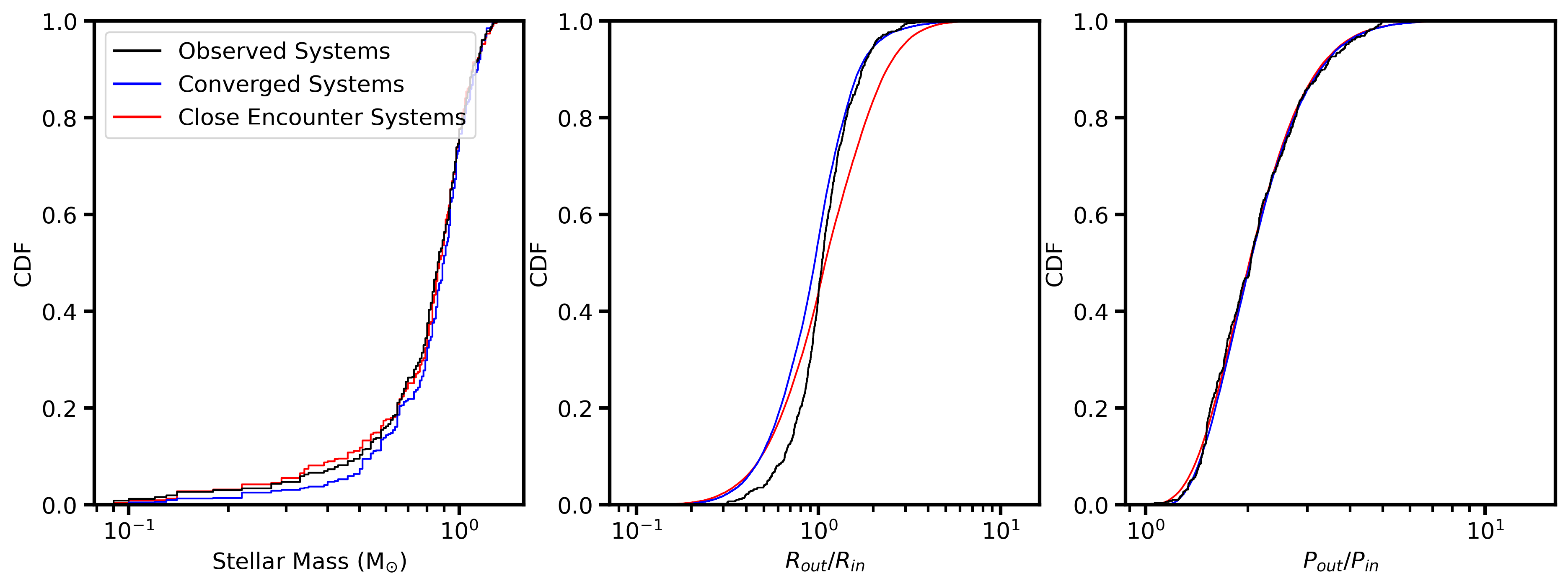} 
\caption{Cumulative distribution functions (CDFs) of stellar mass, radius ratio and period ratio between neighboring planets. The black curves show the $239$ confirmed multi-planet ($N_p\geq 3$) systems from the NASA Exoplanet Archive. In blue and red, we display the corresponding values from our simulations. Note that the period ratios are initial period ratios. The red curves denote systems that experienced some close encounters and are hence discarded (see Section \ref{sec: 2.4}). The blue curves represent the systems that completed disk migration.}
\label{fig: 1} 
\end{center} 
\end{figure*}

We focus on Type-I migration of low-mass planets that do not carve a gap in protoplanetary disks \citep[see][and references therein]{Kley_2012}. Our approach is to use N-Body simulations with a simple prescription for Type-I migration based on the results of hydrodynamic simulations. It should be noted that several studies have characterized the formation of MMRs using hydrodynamic simulations directly \citep{Cresswell, ataiee2021pushing, mcnally2019migrating}. Specifically, we investigate the stage of disk migration after planets have grown to their final masses (post-disk, mass loss and mergers may still be possible but are not simulated here). See also the alternative approach by e.g. \citet{Izidoro} who grow the planets while migrating them. One benefit of our methodology is that we have better control of the final orbital architecture: stellar mass, planet mass, intra-system uniformity, etc. As will be described shortly, we strive to reproduce observed Kepler-like planets.

Type-I migration torque depends on local disk conditions, including the surface density $\Sigma$ and the disk aspect ratio $h\equiv H/R$. To simplify our model, both were assumed to be power laws:

\begin{equation}\label{eq: 1}
    \Sigma = \Sigma_{\rm 1AU} \left( \frac{r}{1AU} \right)^{-\alpha} 
\end{equation}

\begin{equation}\label{eq: 2}
    h = h_{\rm 1AU} \left( \frac{r}{1AU} \right)^\beta
\end{equation}

\noindent where $\Sigma_{\rm 1AU}$ and $h_{\rm 1AU}$ are the surface density and the aspect ratio at $1$ AU and $r$ is the radial distance from the host. $\alpha$ and $\beta$ are the power-law indices. In this work, we set $\alpha$ to $1.5$ following the MMSN \citep{Hayashi}, and the disk flaring index $\beta$ to $0$ i.e. no flaring. We drew $\Sigma_{\rm 1AU}$ from a log-uniform distribution between $10$ and $10,000\text{ g~cm}^{-2}$. We adopt a wide range for the surface density to encompass the MMSN \citep[$\sim1700\text{ g~cm}^{-2}$][]{Hayashi} as well the Minimum-Mass Extrasolar Nebula which is estimated to be a factor of few denser than MMSN \citep{ChiangMMEN,Dai_MMEN,He_MMEN}. Additionally, sub-Neptune formation may occur in transitional disks \citep{Lee2016transitional} or in truncated disks \citep{Dupuy444}, both of which may have significantly depleted surface densities.

Our prescription for migration timescales follows \citet{Pichierri2018} and references therein. $\tau_a$ (Equation \ref{eq: 3}) is the decay timescale for the semi-major axis, and $\tau_e$ (Equation \ref{eq: 4}) is the timescale for eccentricity damping by the disk. The ratio of $\tau_a$ to $\tau_e$ is the $K$-factor (Equation \ref{eq: 5}):  

\begin{equation}\label{eq: 3}
    \tau_a \simeq \frac{1}{2.7+1.1\alpha}\frac{M_*}{m}\frac{M_*}{\Sigma a^2}\frac{h^2}{\sqrt{GM_*/a^3}}
\end{equation}

\begin{equation}\label{eq: 4}
    \tau_e \simeq \frac{1}{0.780}\frac{M_*}{m}\frac{M_*}{\Sigma a^2}\frac{h^4}{\sqrt{GM_*/a^3}}
\end{equation}

\begin{equation}\label{eq: 5}
    K = \frac{\tau_a}{\tau_e} \simeq \frac{0.780}{2.7+1.1\alpha}h^{-2}
\end{equation}

\noindent where $M_*$ is the stellar mass, $m$ is the planet mass, and $a$ is the planet's semi-major axis. We drew $K$ from a log-uniform distribution between $10$ and $1000$.  The aspect ratio $h$ is derived by inverting Equation \ref{eq: 5} and is generally between $0.01$-$0.1$.

To prevent the planets from falling onto the host star, we introduced an inner disk edge where the migration torque is reversed \citep{Masset_2006,Izidoro,Wong2024}. Resonant capture occurs only during convergent migration \citep[where period ratio decreases, see e.g.][]{Batygin_resonance}. The inner disk edge further facilitates the formation of resonant chains by converting divergent encounters into convergent ones, allowing longer-period planets to catch up with the planets that have reached and stopped at the inner disk. We fixed the inner disk edge at $0.05$ AU, with a width of $0.01$ AU to represent the location of magnetospheric truncation where the magnetic forces from the host star disrupt the accretion flow (e.g.~\citealt{Shapiro1983}). Following the suggestion of \citet{Batygin_disk_edge} that the period associated with a test particle at the inner disk edge depends weakly on host star masses, we used the same inner disk edge in our simulations.

In our simple prescription of Type-I migration with an inner disk edge, the innermost planet is solely responsible for halting the chain's inward migration. This is because the net migration torque is reversed in a narrow region of 0.01 AU centered at 0.05 AU where typically only one planet can reside. We found that in $\sim10\%$ of our simulations, the innermost planet is pushed by longer-period planets through the disk inner edge. Lower-mass innermost planets (median mass and 68\% middle quantile range is $10^{0.2\pm0.4}M_\oplus$) are particularly susceptible to being pushed through the disk edge.

All of these disk migration prescriptions were implemented  with the {\tt type\_I\_migration} \citep{kajtazi2023mean} scheme in {\tt REBOUNDx} \citep{Tamayo_x} and {\tt REBOUND} \citep{Rein}. We used the symplectic {\tt WHFAST} integrator \citep{1991AJ....102.1528W, Rein2015}. The time step was set to $1/20$ of the Keplerian orbit at the inner disk edge. {\tt REBOUNDx} combines N-Body simulations with migration parameterizations based on hydrodynamic simulations.

Our systems were generally evolved for $3\tau_a$. Protoplanetary disk lifetime depends on stellar mass \citep{ribas2015protoplanetary}, with some disks around low mass stars lasting $20$ Myr or longer \citep{long2025first, silverberg2020peter}. Disks around stars up to $2 M_\odot$ may survive up to $15$ Myr \citep{wilhelm2022exploring}, though typical disk lifetimes are $\sim3$ Myr \citep{haisch2001disk, li2016lifetimes}. Based on these disk dispersal constraints, we set an upper limit on the integration time of $10$ Myr. We also set a minimum integration time of $30$ kyr if $3 \tau_a < 30$ kyr.

\subsection{Mimicking Observed Kepler-like Systems} 
\label{sec: 2.2}

In our simulations, we strive to match the stellar masses, planetary radii/masses, orbital period ratios, multiplicity, and intra-system uniformity observed in confirmed exoplanetary systems. We downloaded the confirmed planets from NASA Exoplanet Archive \footnote{\url https://exoplanetarchive.ipac.caltech.edu} on August first 2024, focusing on systems with at least three transiting planets. We excluded planets that are likely too massive for Type-I migration ($> 30M_\oplus$). This resulted in a total of $239$ confirmed multi-planet systems with up to seven planets. 

\begin{figure}[ht] 
\begin{center} 
\includegraphics[width = 1.0 \columnwidth]{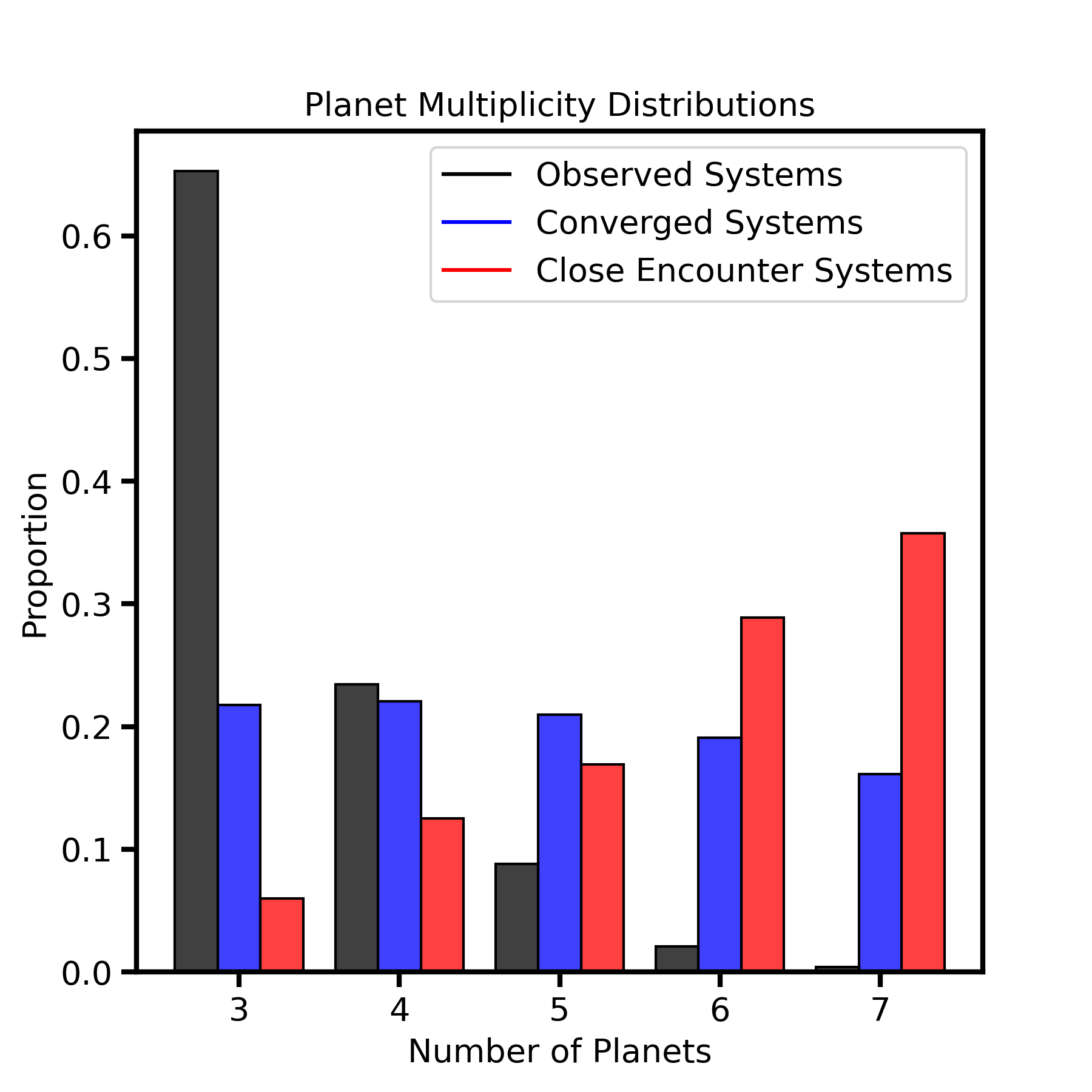} 
\caption{Planet multiplicities for observed systems (black), systems that experienced close encounters (red), and systems that complete disk migration with no close encounters (blue). Note that observed multiplicity is smaller than the actual multiplicity. Section \ref{sec: 2.4} explains why close encounters are more common in higher-multiplicity systems. The blue systems were used in our subsequent analysis.} 
\label{fig: 2} 
\end{center} 
\end{figure}

To simulate a planetary system, we first drew the name of a confirmed planetary system. We adopted the reported stellar mass, and we used the observed radii of the planets while preserving their order of semi-major axes. The innermost planet was initialized at $0.1$ AU i.e. substantially far away from the inner disk edge of 0.05AU. Again, we aimed to simulate the final assembly of a resonant chain just before the planets reach the disk inner edge. The initial orbital periods of longer-period planets were set to be non-resonant. Specifically, we place additional planets by drawing orbital period ratios from a natural lognormal fit (16th percentile: $1.54$, 50th percentile: $2.03$, 84th percentile: $2.88$) of the observed period ratio distribution between neighboring planets \citep[e.g.][]{Fabrycky2014,Weiss2018}. The vast majority of neighboring planets had initial period ratios between 1.2-5 (see Fig. \ref{fig: 1}). 

\begin{figure*} 
\begin{center} 
\includegraphics[width = 2.\columnwidth]{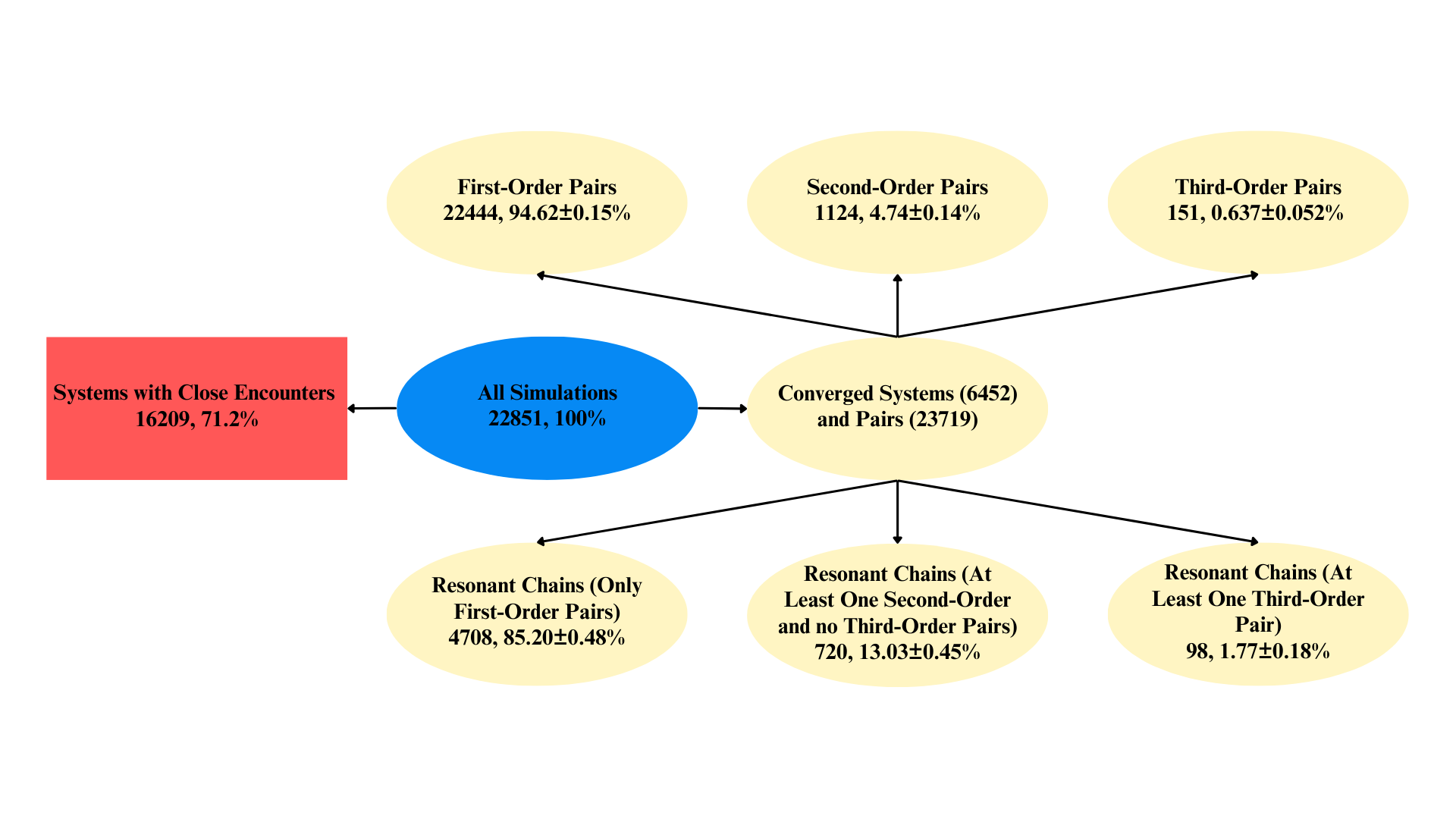} 
\caption{More than 70\% of our simulations were discarded due to having close encounters. Among the remaining systems: $\sim 5\%$ and $\sim 0.5\%$ of resonant planet pairs were in librating second- or third-order resonances; $\sim 13\%$ and $\sim 1\%$ of planetary systems classified as three-body resonant chains and two-body resonant chains contain at least one second-order or third-order resonance.} 
\label{fig: 3} 
\end{center} 
\end{figure*}

The actual multiplicity of a confirmed planetary system is probably higher than the observed multiplicity \citep[see e.g.][]{Zhu_Dong, turtelboom2024searching}. As the observed sample is dominated by three planet systems, we generated more four to seven planet systems until all multiplicities had roughly the same number of systems (Fig. \ref{fig: 2}). We add more planets by repeatedly drawing from the same logarithmic distribution of period ratios as we described above. Moreover, to capture the intra-system uniformity in planet size \citep['peas-in-a-pod' pattern][]{Weiss_peas,Wang_uniform,Millholland_peas}, we also fitted a natural lognormal distribution of radius ratios between neighboring planets (16th percentile: $0.708$, 50th percentile: $1.06$, 84th percentile: $1.57$). Each new planet has a radius that depends on the radius of the planet directly interior to it and a random sample from this log-normal distribution.

Planet masses were determined from the radii and the mass-radius relationship {\tt Forecaster} \citep{Chen2017}. In essence, this is a power law: $R = M^{0.279}$ for $M<2 M_\oplus$ and $R = M^{0.59}$ for $M>2 M_\oplus$) with intrinsic scatter. Sometimes, {\tt Forecaster} can produce masses that are perhaps so large that Type-I migration is not an accurate description. If we drew a mass $>30 M_\oplus$, we simply drew again.

In Figures \ref{fig: 1} and \ref{fig: 2}, we compare our simulated systems with observation. Our simulated sample emulates the observed sample in terms of stellar mass and neighboring planet radius and period ratios. In other words, our simulated planetary systems retain the previously reported stellar-mass-planet-size correlation \citep[e.g.][]{Wu2019_mass} and the `peas-in-a-pod' pattern \citep{Weiss_peas,Wang_uniform,Millholland}.

Although some studies have prescribed nonzero initial eccentricity and inclinations \citep[e.g.][]{Izidoro}, we set the initial eccentricity and orbital inclination of each planet to zero since we expect disk damping prior to resonant encounters. The mean anomaly was drawn from uniform distributions between 0 and $360^\circ$. 

\subsection{Identifying Mean-Motion Resonance} 
\label{sec: 2.3}

\begin{figure*} 
\begin{center} 
\includegraphics[width = 2.\columnwidth]{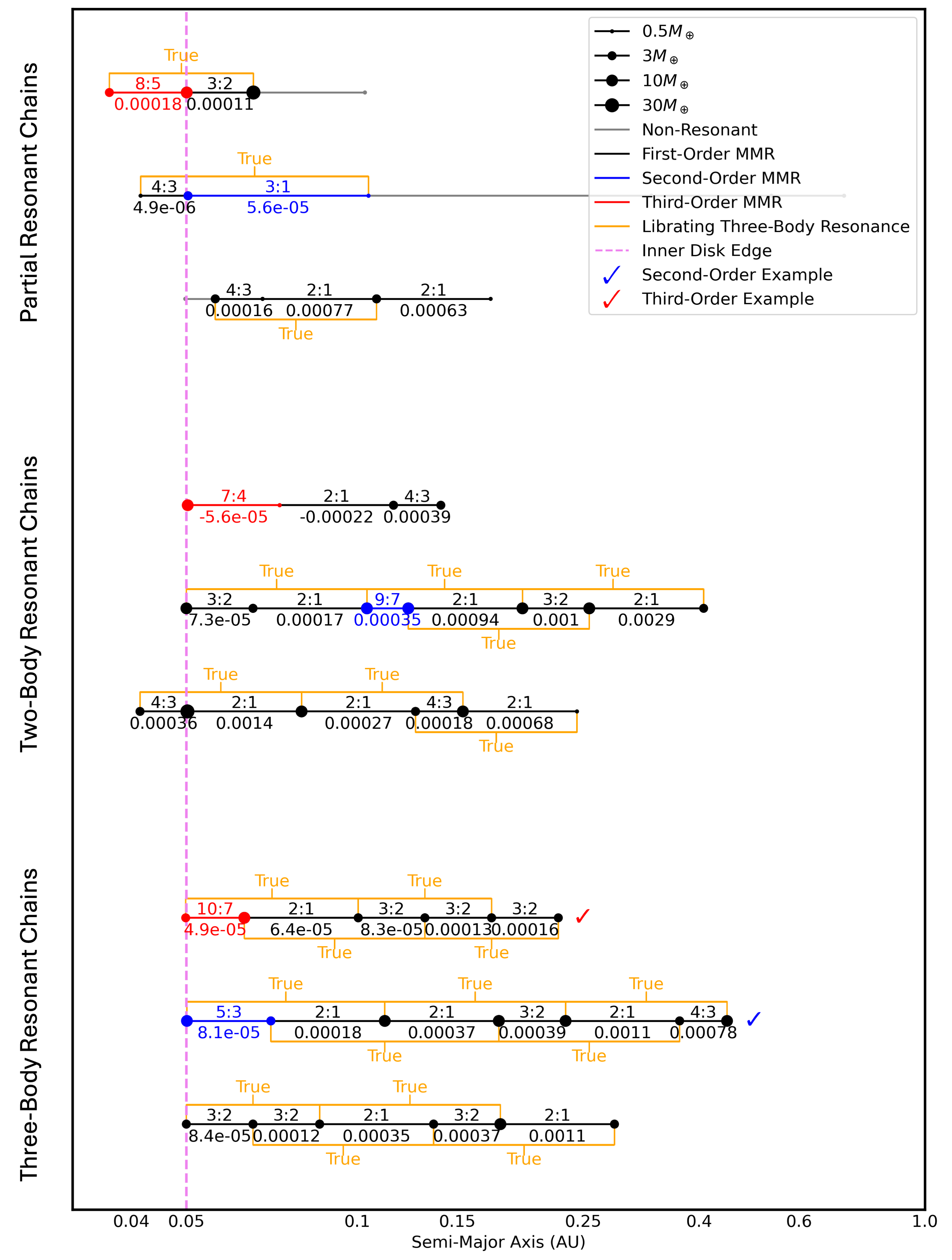} 
\caption{A representative set of nine systems at the end of our simulations. We label each resonant pair of planets with its integer ratio and deviation $\Delta$. Colors indicate the order of resonance: nonresonant in grey, first-order in black, second-order in blue, and third-order in red. Librating three-body resonant angles are shown in orange. Two case study systems in Section \ref{sec: 3} are indicated with check marks. The top three systems are `Partial Resonant Chains' (at least one pair not in MMR); the middle three are `Two-Body Resonant Chains'; the bottom three are `Three-Body Resonant Chains.' In our simulations, low-mass planets occasionally pass through the disk's inner edge (see Section \ref{sec: 2.1}).}
\label{fig: 4} 
\end{center} 
\end{figure*}

In the literature, $\Delta$ is frequently used as a convenient metric to identify MMR: 

\begin{equation}\label{eq: 6} 
    \Delta \equiv \frac{P_c/P_b}{p/q} - 1
\end{equation}

\noindent for a pair of planets bc near the integer ratio $p$:$q$. When a planetary system's dynamical state is unknown, measuring $\Delta \simeq 0$ quickly identifies near-resonant planets \citep[e.g.][]{huang2023and,dai2024prevalence}. However, the hallmark of true resonance is the libration of a resonant angle in the presence of a separatrix, a generalized coordinate for the resonant Hamiltonian \citep{Murray}. The two-body resonant angle, $\phi_{bc}$, is given by Equation \ref{eq: 7} where $\lambda$ are the planets' mean longitudes and $\hat{\varpi}_{bc}$ (Equation \ref{eq: 8}) where $e_b$, $e_c$ are the eccentricities and $f$, $g$ are coefficients from the expansion of the disturbing function \citep{Sessin,Henrard1986, Wisdom1986, Batygin_resonance}.

\begin{equation}\label{eq: 7}
    \phi_{\rm bc} = q\lambda_{b} - p\lambda_{c} + (p-q)\hat{\varpi}_{bc} 
\end{equation}

\begin{equation}\label{eq: 8}
    \hat{\varpi}_{bc} = \arctan{\left( \frac{fe_b\sin{\varpi_b} + ge_c\sin{\varpi_c}}{fe_b\cos{\varpi_b} + ge_c\cos{\varpi_c}} \right)} 
\end{equation}

\citet{Hadden_resonance} demonstrated that the mixed longitude of the pericenter, $\hat{\varpi}_{bc}$, is an acceptable approximation for higher-order. We select the mixed angle as it reduces the two-body resonance to a single degree of freedom. If the two-body resonance is in libration, $\phi_{bc}$ will librate with the mixed angle included. We estimate the libration amplitude following \citet{Millholland_2018_resonance}:

\begin{equation}\label{eq: 9}
    A = \sqrt{\frac{2}{N} \sum (\phi - \langle \phi \rangle)^2} 
\end{equation}

\noindent where $N$ is the number of snapshots, $\phi$ is the resonant angle, and $\langle \phi \rangle$ is the mean resonant angle over the $N$ samples (a proxy for the equilibrium point). We identify MMR as pairs of planets with libration amplitude $A<90^\circ$ at the end of our disk migration simulation.  

A triplet of planets can also be engaged in a zeroth-order three-body, Laplace-like MMR e.g. the TRAPPIST-1 planets \citep{Agol2021} or our Galilean moons \citep{Peale1976}. Higher-order three-body resonances may play a role in resonance capture and stability \citep{petit2021integrable}, but, as the only known case is the first-order three-body resonance of K2 138efg \citep{cerioni2023six}, we concentrate on zeroth-order three-body resonances in this work.

We construct the three-body resonance angle $\phi_{bcd}$ following Equation \ref{eq: 10}.


\begin{align}\label{eq: 10}
    \phi_{bcd} &= 
    (p_{cd}-q_{cd})(q_{bc}\lambda_b - p_{bc}\lambda_c) \notag\\
    &\quad + (p_{bc}-q_{bc})(-q_{cd}\lambda_c + p_{cd\lambda_d)} 
\end{align}

\noindent where $p$ and $q$ are the integers of two-body MMR for the $b$-$c$ pair and $c$-$d$ pair. 

To ensure that the libration of three-body MMR prefers a center of 180$^\circ$, we follow the suggestion of \citet{Siegel2021} and do not reduce Equation \ref{eq: 10} by the greatest common denominator. We again identify three-body MMR as triplets with libration amplitude $A<90^\circ$. At the end of disk migration simulations, we examined the prevalence of both two-body and three-body MMR.

\begin{figure*}
\begin{center} 
\includegraphics[width = 2.\columnwidth]{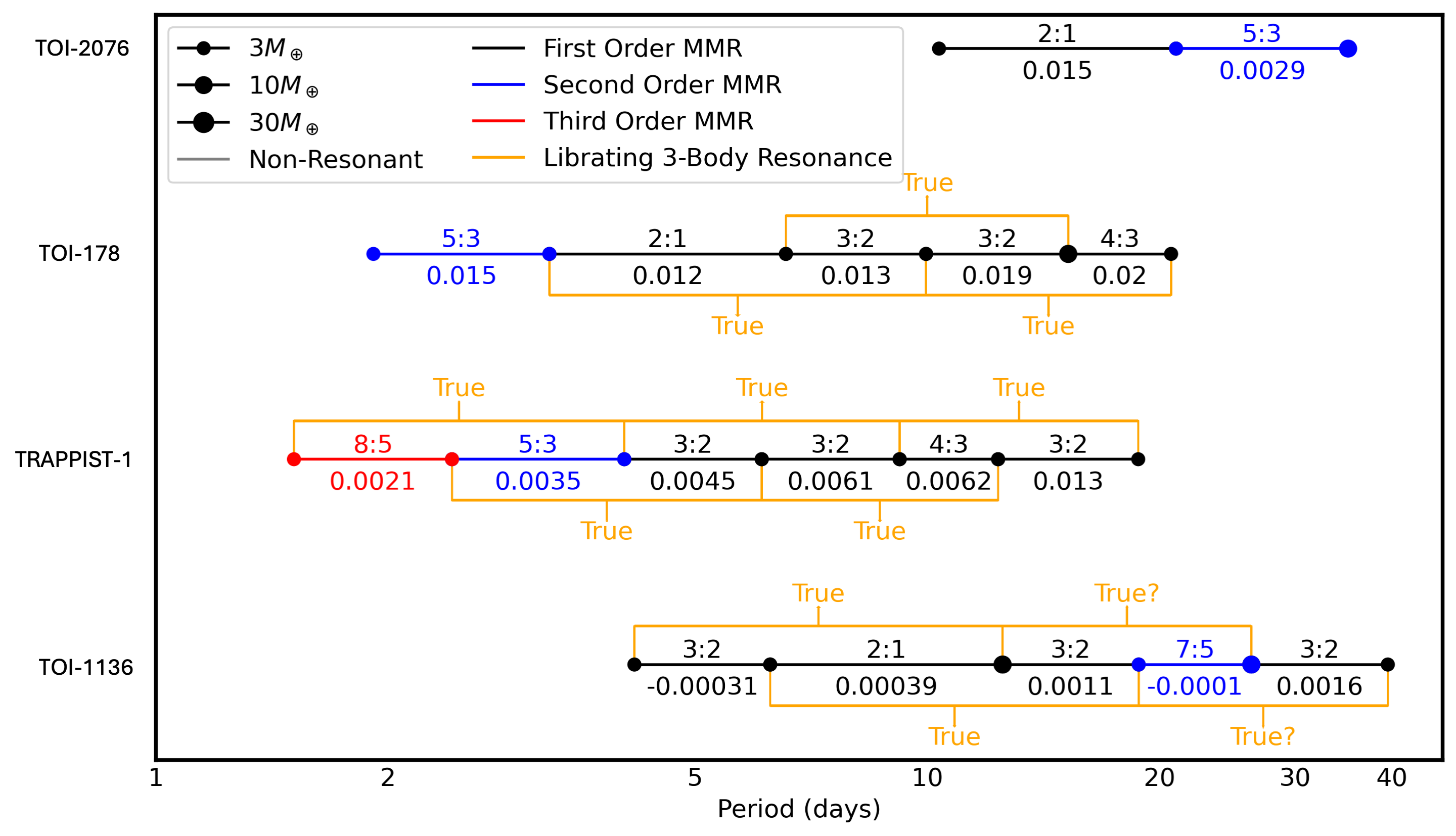}
\caption{A subset of observed resonant chains plotted like the simulated systems shown in Fig. \ref{fig: 4}. Unlike Fig. \ref{fig: 4}, these observed systems are displayed with respect to period to reduce whitespace. In most example observed systems, it is unclear if the two-body resonant angles librate. We choose to mark the most proximal MMRs. Librating triplets are indicated in accordance with observation.}
\label{fig: 5} 
\end{center} 
\end{figure*}

If there are more than two or three planets in a resonant arrangement in a system, we can prescribe an N-body resonant angle where N is the number of planets in the arrangement. However, these are very difficult to constrain even in systems that are deeply in resonance like TOI-1136 \citep{dai2023toi}. Instead, it is appropriate to define a ``resonant chain'' where all planets pairs and/or triplets in the system have librating two- and/or three-body resonant angles. We adopt the following terms to describe resonant chains (see also examples in Fig. \ref{fig: 4}):

\begin{enumerate} 
    \item Complete Three-Body Resonant Chain: All neighboring triplets of planets exhibit librating three-body resonant angles and all neighboring pairs have librating two-body resonant angles. 
    \item Complete Two-Body Resonant Chain: All neighboring pairs exhibit librating two-body resonant angles and only some (or none) of the triplets have librating three-body resonant angles. 
    \item Partial Resonant Chain: Some planet pairs exhibit librating two-body resonant angles while others are non-resonant. Any number of triplets may have librating three-body resonant angles.
\end{enumerate}

In all cases, the two-body MMRs can be of any order. We examined all 36 first-, second-, and third-order MMRs ranging from period ratio of $1.1$ ($11$:$10$) to $4$ ($4$:$1$). This range encompasses the smallest observed pairwise period ratios \citep[Kepler-36bc 7:6, see][]{carter2012kepler} to the widest third-order resonance, 4:1. We compiled a library of corresponding $f$ and $g$ coefficients for each $p$ and $q$ using the {\tt disturbing\_function.get\_fg\_coefficients} routine from {\tt celmech} \citep{hadden2022celmech}.

\subsection{Handling Close Encounters} 
\label{sec: 2.4}

We ran a total of $22,851$ simulations, roughly $\sim70\%$ of which experienced close encounters. Since symplectic integrators are not designed to handle close encounters \citep{1991AJ....102.1528W}, we stopped and discarded any simulations where planets ventured within five mutual Hill radii of each other, following \citet{Weiss2018} who noted that Kepler systems tend to have significant spacings between planets. We note that close encounters do not necessarily prevent a system from developing a resonant chain eventually \citep{Izidoro}. However, our simulation setup is currently not equipped to accurately predict the outcome of close encounters. We defer that to a future work as we are more concerned with determining the orders of MMRs that form during disk migration than running comprehensive population synthesis and planet growth calculations.

As evident in Fig. \ref{fig: 2} and \ref{fig: 10}, systems with close encounters generally feature higher planet multiplicity, more massive outer planets, and faster migration than the systems that quiescently completed disk migration. Close encounters are known to be more common in high multiplicity systems \citep{smith2009orbital} and it has been suggested that longer resonant chains are less stable as they tend to experience a secondary resonance between a libration frequency and a fraction of the synodic frequency \citep{Matsumoto2012,Pichierri2020,Goldberg_stability}. A more massive outer planet also makes resonances less stable and more prone to disruption \citep[e.g.][]{Goldreich2014,Deck_Batygin,Xu0217}. Finally, faster migration makes it less likely for planets to capture into resonance \citep{Batygin_resonance}. The fact that we initialized systems with zero mutual inclination may have further increased close encounters.

\section{Case Studies of Systems that Developed Higher-Order Resonance}
\label{sec: 3}

\begin{figure*}[h]
\begin{center}
\includegraphics[width = 0.8\linewidth]{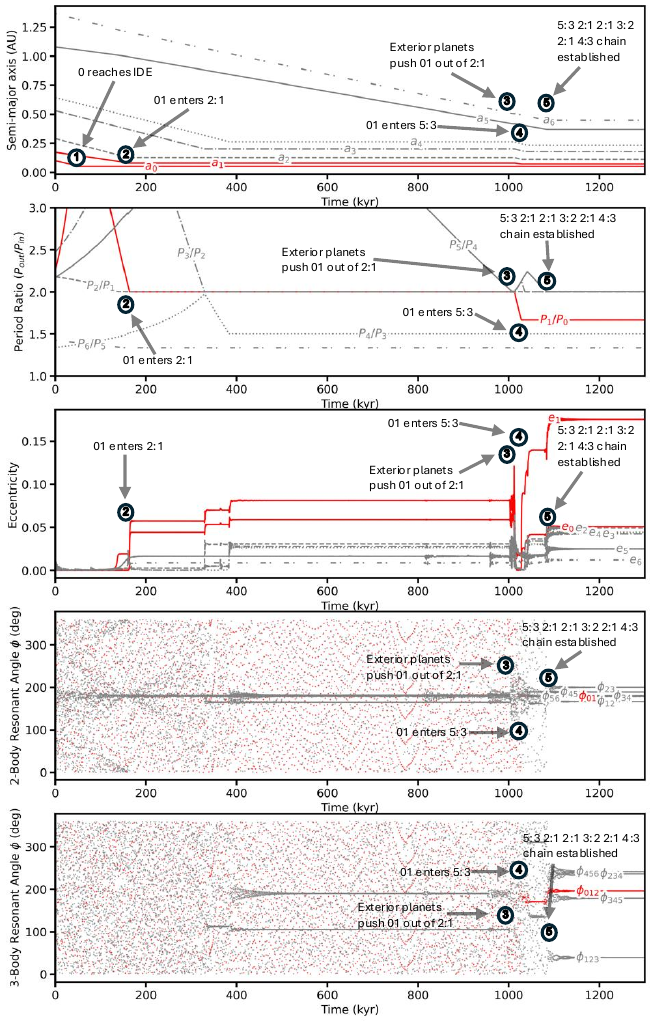}
\caption{The migration history of a resonant chain with a second-order resonance (innermost two planets 0 and 1, shown in red). Planet 0 and 1 were initially captured into a 2:1 resonance. However, as longer-period planets joined the resonant chain, the 2:1 resonance broke and the planets captured into the nearby 5:3 resonance. A few key milestones of the evolution have been labeled. See Section \ref{sec: 3} for a detailed description of this system. Tab. \ref{tbl: 2} contains the system parameters.}
\label{fig: 6}
\end{center}
\end{figure*}

Both second- and third-order MMR emerged in our disk migration simulations. Before we present population-level results (Section \ref{sec: 4}), let's examine the migration history (Fig. \ref{fig: 6}) of a particular system that ended up with a 5:3 second-order resonance. The system's architecture is shown in Fig. \ref{fig: 4} (eighth row), and its initial conditions are presented in Tab. \ref{tbl: 2}. 

In this particular system, seven planets were initialized between $0.1$ and $1.37$ AU around a $0.89 M_\odot$ host star in a disk with an aspect ratio of $0.0395$ and a low surface density of $16.6$ g/cm$^2$.

The innermost two planets are the pair that eventually captured into a second-order 5:3 MMR. We denote these two planets 0 and 1 and highlight them in red in Fig. \ref{fig: 6}. Planet 0 is more massive than planet 1: 8.0$M_\oplus$ v.s. 4.8$M_\oplus$. Therefore, planet 0 initially migrated faster than planet 1 i.e. two planets initially experienced divergent migration. 

A key milestone of their evolution is marked (1) in Fig. \ref{fig: 6}. When planet 0 reached the inner disk edge (around 50 kyr), its migration is effectively stopped, allowing planet 1 to catch up. This highlights how the inner disk edge converts divergent migration to convergent migration.

At milestone (2) or 160 kyr, planets 0 and 1 captured into a 2:1 MMR. They remained in this resonance until milestone (3) near 1010 kyr. In the intervening time 160-1010 kyr, longer period planets successively arrived at the inner edge and formed a chain of first-order resonances from planet 0 to 4.

\begin{figure*}
    \begin{center}
    \includegraphics[width = 2.\columnwidth]{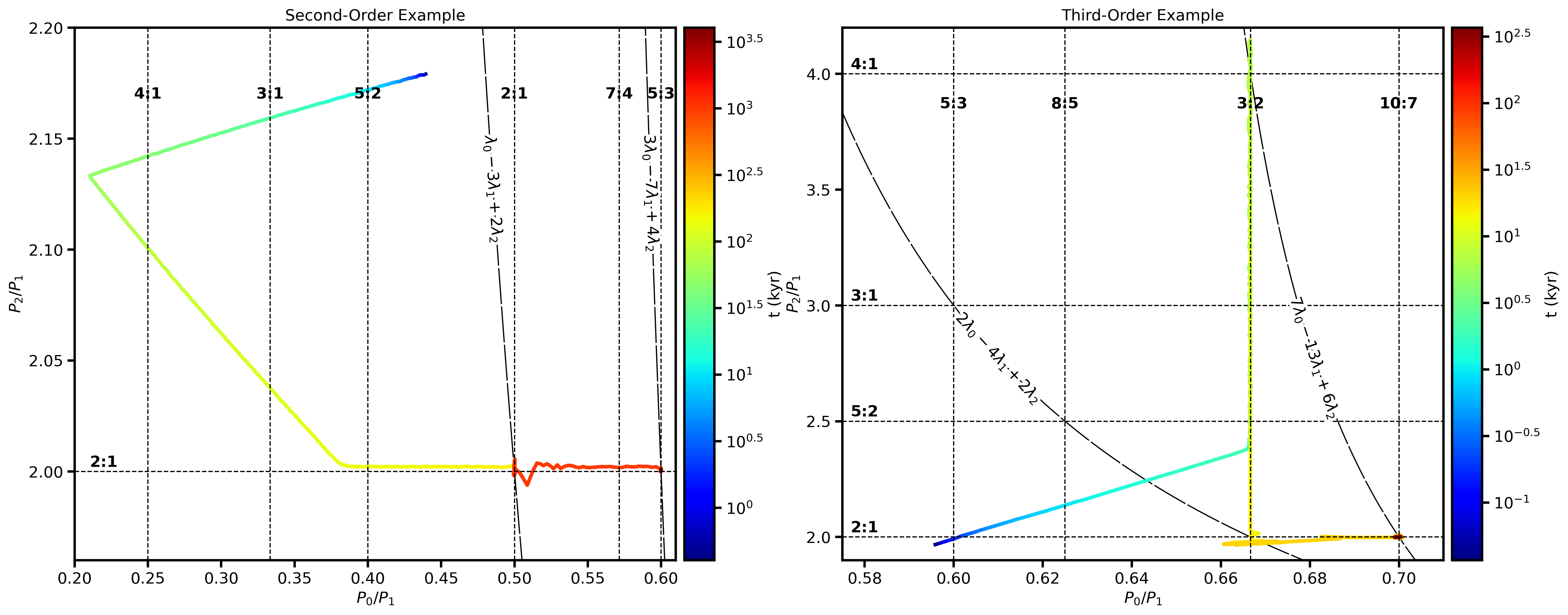}
    \caption{The period ratio evolution of a triplet of planets that end up in higher-order MMR (case study systems presented in Section \ref{sec: 3}). The color coding indicates the time since the start of the simulation. Horizontal and vertical straight lines indicate two-body resonance between the inner and outer pair of planets. Dashed curves (Equation \ref{eq: 11}) show the evolution track if a librating three-body Laplace-like resonance is preserved. In both systems, the planets evolved horizontally i.e. they captured into 5:3 and 10:7 resonance directly without forming a three-body resonance first.}
    \label{fig: 7}
    \end{center}
\end{figure*}

At milestone (3), planets 5 and 6 joined the inner resonant chain of planets 0-4. It appears that as planets 5 and 6 tried to join the resonant chain, a wave of perturbations were sent through the existing chain. As a result, the innermost pair 01 left the 2:1 MMR as their periods continued to decrease.

Before reaching the next first-order MMR, 3:2, the innermost pair 01 captured into the second-order MMR 5:3 at milestone (4) around 1030 kyr. Thanks to the original 2:1 MMR, planet 0 and 1 had low but non-zero orbital eccentricity of order 0.01 before encountering the 5:3 resonance. The non-zero eccentricity was critical for strengthening the second-order resonant interaction (e$^{|p-q|}$) and facilitated the capture, see Sections \ref{sec: 4.1} and \ref{sec: 4.6} for more discussion.

At milestone (5) or 1080 kyr, all seven planets fully captured into a resonant chain, all constituent two-body and three-body resonant angles entered a state of libration through to the end of the simulation.

It is worth noting that the initial period ratio of the innermost pair ($P_{out}/P_{in}>2$) was not close to that of the second-order resonance it captured into. Instead, the successful formation of a second-order resonance in this system is owed to several fortuitous factors:
\begin{itemize}
    \item The low surface density at 1 AU of $16.6$ g/cm$^2$ ($525g/cm^2$ at 0.1AU) resulted in a slow migration rate.
    \item The longer-period planets in this seven-planet system eventually pushed the innermost pair out of the initial 2:1 resonance.
    \item 5:3 is the first strong resonance encountered after the innermost pair left 2:1 resonance before they can reach the next first-order, 3:2.
    \item Due to the initial 2:1 resonance, the innermost pair of planets already had low but nonzero eccentricity before they encountered 5:3 MMR. The eccentricity might have amplified the second-order resonance.
\end{itemize}

Are these conditions difficult to achieve in typical Type-I migration simulations? We discuss that in the next section. In the Appendix, we also show the formation of a third-order 10:7 resonance in Tab. \ref{tbl: 3} and Fig. \ref{fig: 14}. Its final orbital architecture is also displayed in Fig. \ref{fig: 4}. Very briefly, this is a six-planet system around a 1.0 $M_\odot$ host star. The surface density is close to the MMSN at $2080$ g/cm$^2$. The aspect ratio is $0.0991$. The story is very similar to the one described above, a pair of planets were initially engaged in a 3:2 first-order MMR. However, upon breaking this resonance, the pair captured into the closest third-order resonance, 10:7, before encountering any second- or first-order resonances. The initial first-order MMR may be important for setting up the capture into third-order MMR because the first-order MMR produced a small but non-zero eccentricity (see Fig. \ref{fig: 14}).

\section{Population-Level Results}
\label{sec: 4}

On a population level, we found that Type-I disk migration coupled with an inner disk edge can produce higher-order MMRs. While this qualitative result holds, we caution that the quantitative interpretations presented in this section depend strongly on our model assumptions. Most importantly, the assumed range of disk surface densities sets the overall migration rate, which influences the fraction of different orders of MMR.

\subsection{Higher-Order Two-Body MMR Do Not Need to Form in a Three-Body Resonance}
\label{sec: 4.1}

It has been suggested in the literature that higher-order commensuralities may emerge through the help of three-body Laplace-like resonances. For instance, \citet{Huang_Ormel} and \citet{pichierri2024formation} postulated that the inner three planets of the TRAPPIST-1 system could have initially formed in a three-body Laplace-like resonance comprised of two first-order two-body resonances, 3:2-3:2. Subsequent divergent migration may have moved the planets out of the 3:2 resonances but preserved the three-body resonance, until the encounter with another three-body resonance left bc and cd near their present period ratios of approximately 8:5 and 5:3 (8:5-5:3 and 3:2-3:2 have the same three-body resonant angle, $\phi_{bcd} = 2\lambda_b-5\lambda_c+3\lambda_d$).

Our simulations, however, suggest that pre-existing three-body resonances are not necessary for higher-order capture. If a particular three-body Laplace resonance is preserved, a system would evolve along the dashed curves, satisfying Equation \ref{eq: 11} \citep{rath2022criterion} in a $P_0/P_1$-$P_2/P_1$ plot (Fig. \ref{fig: 7}). However, this pathway cannot explain the case studies of higher-order resonance presented in Section 3. The migration histories of these two systems are horizontal or vertical i.e. the planets interact under the influence of two-body resonances. In other words, the systems captured into second- and third-order two-body resonances without first establishing a three-body Laplace-like resonance.

\begin{equation} 
\label{eq: 11} 
\frac{P_2}{P_1} = \frac{p_{12}}{{p_{01} + q_{12} - q_{01}(P_0/P_1)^{-1}}}
\end{equation}

\begin{figure*}
    \centering
    \includegraphics[width=\linewidth]{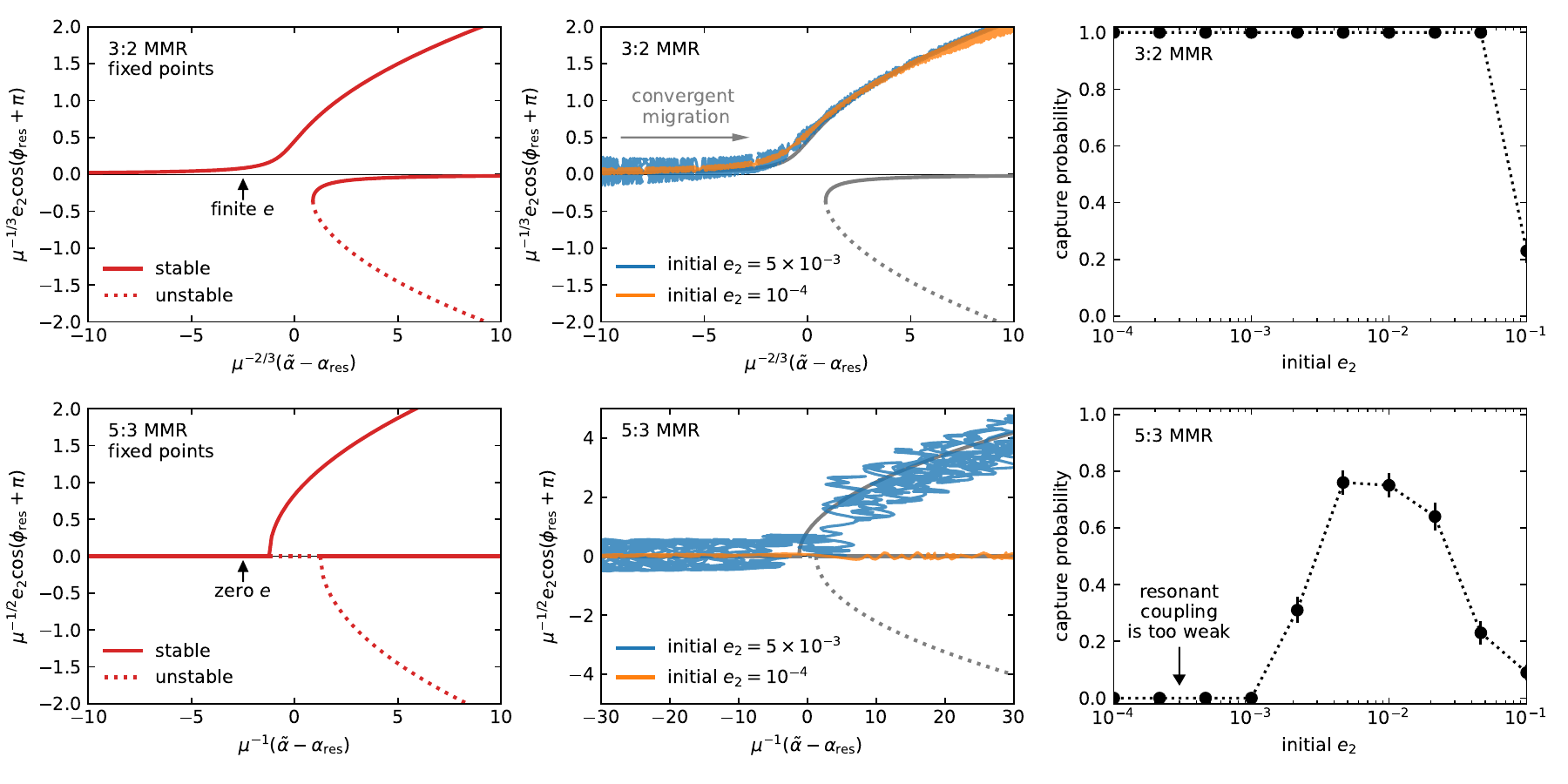}
    \caption{A comparison between capturing into first-order and second-order MMR. For simplicity, we consider a restricted three-body problem where the outer planet is a test particle and the inner planet has planet-to-star mass ratio $\mu$. Left panels: fixed points of the Hamiltonian. Compared to similar plots in the literature (e.g., Fig.~1 in \citealt{Batygin2015}), we opt for explicitly expressing the resonant parameter (horizontal axis) and canonical variable (vertical axis) in terms of physical quantities. The resonant parameter (horizontal axis), with $\alpha_{\rm res} = (q/p)^{2/3}$ and $\tilde\alpha = (a_1/a_2)(1+\frac{p}{p-q}e_2^2)$, is conserved in the absence of migration. For first-order MMR, the eccentricity of the fixed point smoothly increases as the system approaches resonance. However, for second-order MMR, the fixed point stays at $e=0$ until encountering resonance. Center panels: simulations at different initial eccentricities. In all runs, we initialize the pair at $2\%$ away from the resonance, and evolve the massless outer planet with $\tau_{\rm a}=10^7 P_1$, $\tau_{\rm e}=10^5 P_1$. For a second-order MMR, low initial eccentricity causes insufficient coupling at resonance, and the system cannot become captured by following the stable fixed point with finite eccentricity. Right panels: capture probability for the toy problem as a function of initial eccentricity. Each point is estimated using 100 simulations starting at random phases. Second-order MMR differs from first-order MMR in that low eccentricity can prevent capture \citep{Xu0217}. As a result, eccentricity excitation from a previous first-order MMR capture generally promotes capturing into a higher-order MMR.}
    \label{fig: 8}
\end{figure*}

We stress that roughly 79\% of the higher-order resonances in our simulations eventually become part of a librating three-body Laplace resonance with its neighboring planets. However, these three-body Laplace resonances were established as a result of the two-body resonances rather than the other way around (see Fig. \ref{fig: 6} and Fig. \ref{fig: 8}). Our findings suggest that two-body higher-order MMR can form without the help of a three-body Laplace-like resonance.

Forming higher-order MMRs with three-body resonance \citep{Huang_Ormel} and forming them directly (our simulations) may both happen in nature. When formation is aided by three-body resonance, the planets may migrate divergently and the resulting two-body resonant angles generally do not librate. This is consistent with TRAPPIST-1 bcd, where the two-body resonant angles circulate in most of the posterior samples in \citet{Agol2021}. In direct formation, the planets migrate convergently and capturing into a two-body MMR produces a librating resonant angle. This serves as a natural explanation for planets such as TOI-1136 ef, where the two-body angle is likely librating \citep{dai2023toi}.

\subsection{Eccentricity Excitation Caused by Pre-Existing First-Order MMR Promotes Higher-Order Capture}
\label{sec: 4.2}

In our simulations, we observe that most higher-order MMR pairs were formed after the system encounters and escapes from a previous first-order MMR. As second-order is the dominant population of these higher-order pairs, we focus on that configuration in this subsection. We argue that the eccentricity excitation by the previous first-order MMR promotes the subsequent capture into a second-order MMR.

\begin{figure*}
\begin{center}
\includegraphics[width = 2.\columnwidth]{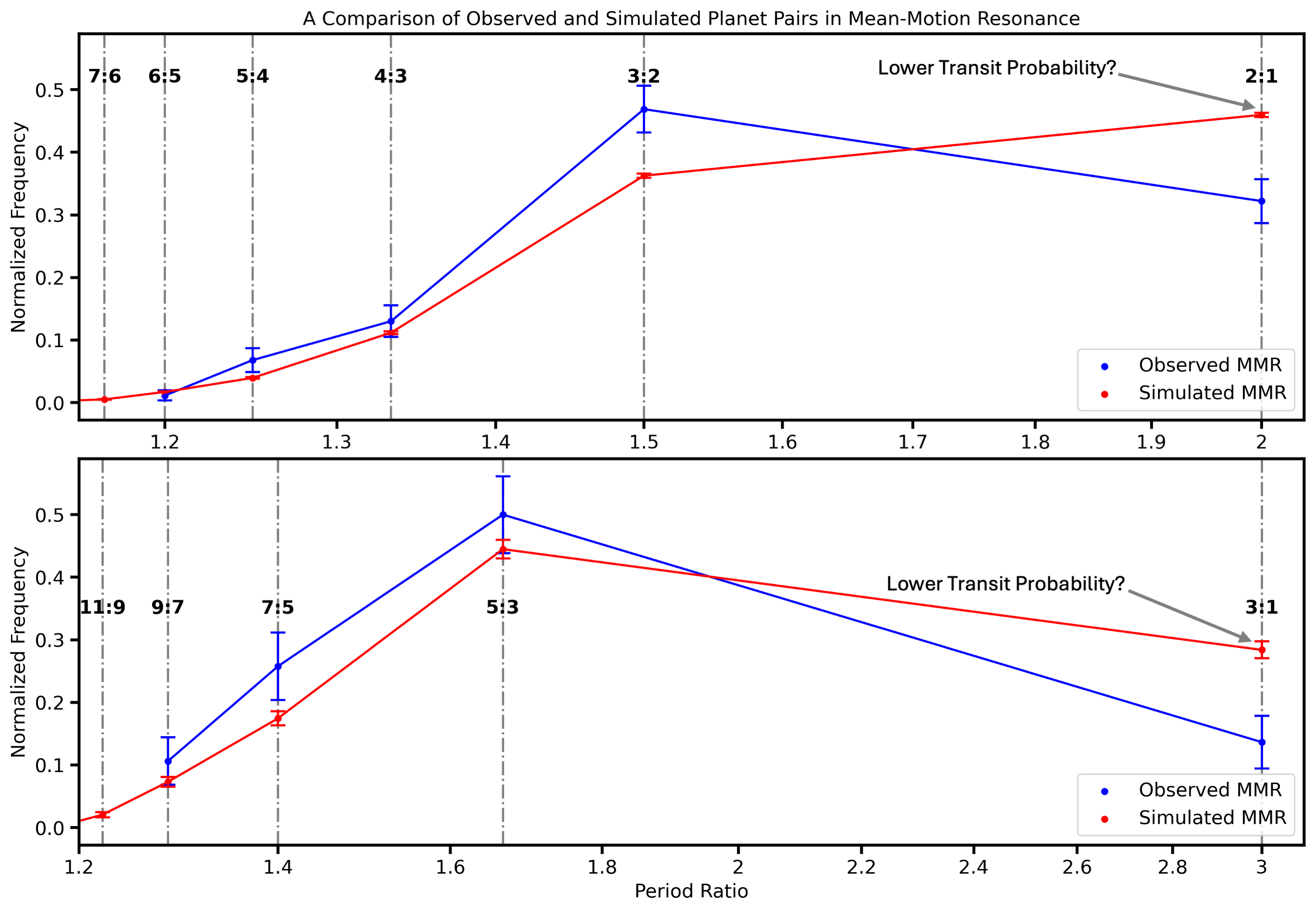}
\caption{The relative frequencies of individual first (top panel) and second-order (bottom panel) MMR in our simulations and in the confirmed all-ages planet sample from \citet{dai2024prevalence}. Note in the simulated samples, all planet pairs have librating resonant angles \citep[the result is practically unchanged if one instead uses the definitions for resonance based on $\Delta$ presented in][]{dai2024prevalence}. In the observed sample, there is not enough information to determine the dynamical state of the planets. Simulations produced more 2:1 and 3:1 resonance than observed. Smaller transit probabilities for inclined planets may underestimate planet counts (and thus over estimate planet separations).}
\label{fig: 9}
\end{center}
\end{figure*}

MMR capture is eccentricity dependent; in the limit of adiabatic evolution (infinitely slow migration), capture is certain at small initial eccentricity, and becomes probabilistic once the eccentricity exceeds $\sim \mu^{1/3}$ for first-order MMR and $\sim \mu^{1/2}$ for second-order MMR \citep{Batygin2015,Xu0217}, where $\mu$ is the planet-star mass ratio. As a result, the claim in Section \ref{sec: 3} that previous eccentricity excitation promotes capture may sound counterintuitive. However, once we consider the finite rate of migration, there is a separate effect that strongly suppresses higher-order MMR capture for small initial eccentricity. This effect originates from a key difference between first-order and higher-order ($|p-q|\geq 2$) MMRs: Since the resonant term in the Hamiltonian is of order $e^{|p-q|}$, for a higher-order resonance, the zero-eccentricity state is a fixed point of the Hamiltonian except for a finite width around the resonance (Fig. \ref{fig: 8} bottom left panel). Therefore, an initially circular orbit will stay at approximately zero eccentricity until the system encounters the resonance. This creates a problem for resonance capture: at zero eccentricity, the strength of resonant coupling also vanishes, and the eccentricity cannot evolve fast enough to follow the stable fixed point of the Hamiltonian to become captured as it does in the adiabatic limit; instead, the system will directly pass through the resonance. The orange line in Fig. \ref{fig: 8} bottom center panel is an example of this; \citet{Xu0217} offers a more quantitative discussion on when capture becomes suppressed by low initial eccentricity. We note that the same problem does not occur for first-order MMR because there the eccentricity of the stable fixed point smoothly increases as the system approaches the resonance. As a result, orbits with low initial eccentricity already acquire finite eccentricity when they encounter resonance and can be easily captured (Fig. \ref{fig: 8} top left and top center panels).

To give a more quantitative estimate on how this mechanism affects capture, we consider a toy problem with a pair of planets where the inner planet has $\mu=3\times 10^{-5}\approx 10M_\oplus/M_\odot$ and the outer planet is a test particle migrating at $\tau_{\rm a}=10^7 P_1$, $\tau_{\rm e}=10^5 P_1$. The mass scale and the migration timescales resemble typical low-mass planets near the edge of the protoplanetary disk. The result is summarized in the right panels in Fig. \ref{fig: 8}. For second-order resonance, capture is suppressed for initial eccentricity $\lesssim 10^{-3}$; meanwhile, for first-order resonance, capture is guaranteed at low eccentricity. (We also see the reduction of capture probability at high eccentricity, which occurs for both first- and second-order MMRs.) The minimum eccentricity required for capturing into a second-order MMR is already larger than what can be produced by the non-resonant interaction between low-mass planets. As a result, capturing into a higher-order resonance requires some additional eccentricity excitation, which can be provided by a previous encounter with a first-order MMR. This pathway offers a promising alternative to the Laplace-like capture described in Section \ref{sec: 4.1}.

This pathway also naturally explains the observation in Section \ref{sec: 4.6} that higher-order MMRs show higher eccentricity. The eccentricity of a captured pair mainly depends on the ratio between eccentricity damping and migration, with $e\sim\sqrt{\tau_e/\tau_{a}}$. To capture a pair into a higher-order MMR through the above pathway, a sufficiently high $\tau_e/\tau_{a}$ is necessary because otherwise the eccentricity inherited from the previous resonance would have been exponentially damped while the pair migrates between resonances.

\begin{figure*}
\begin{center}
\includegraphics[width = 2.\columnwidth]{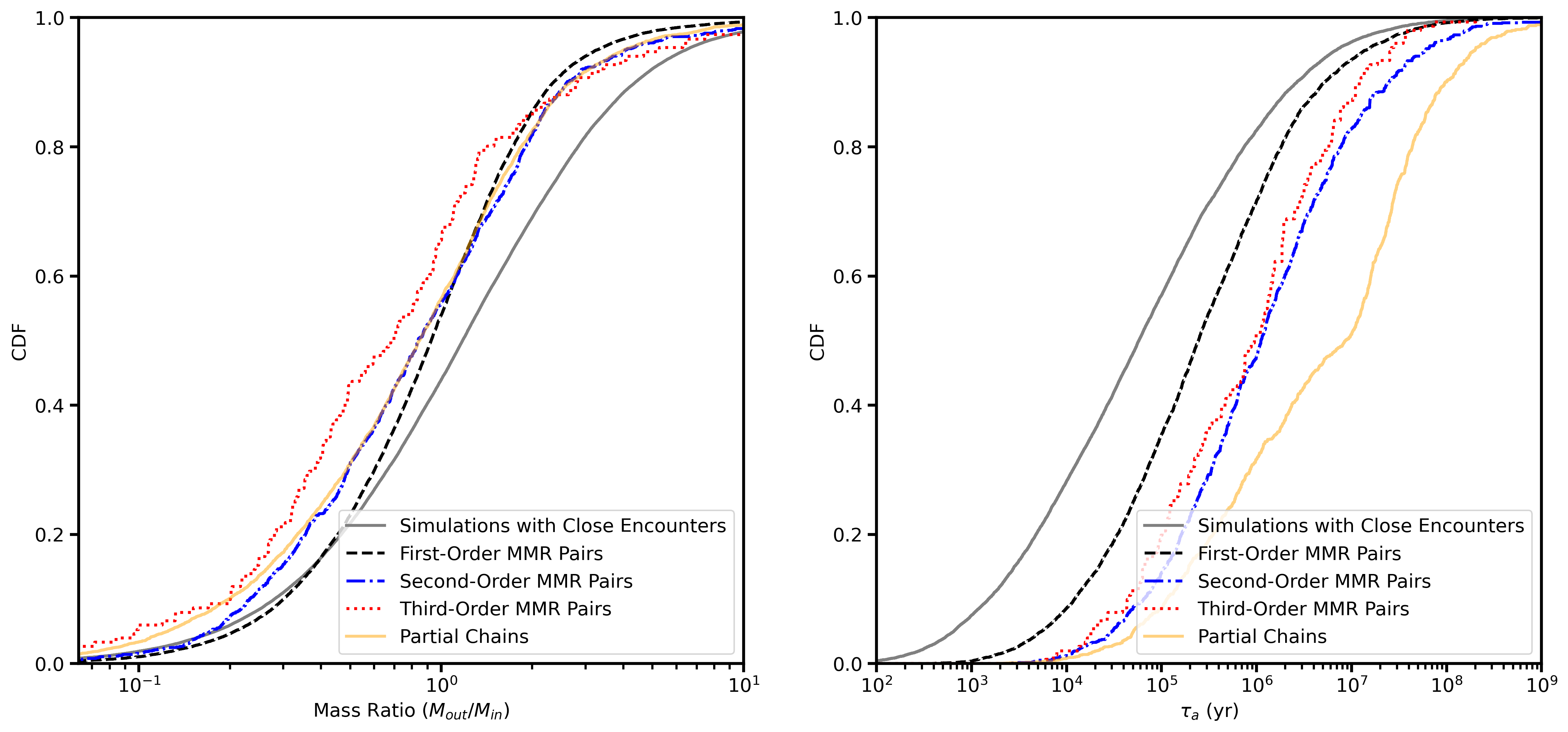}
\caption{CDFs of the mass ratios and the migration timescales $\tau_a$ of the planet pairs in systems that captured into first-, second-, third-order resonance. Systems that experienced close encounters and systems that failed to form complete resonant chains are also shown. Higher-order MMRs preferentially form when the outer planet is less massive and when $\tau_a$ is large, both of which are tied to slow migration.}
\label{fig: 10}
\end{center}
\end{figure*}

\subsection{The Prevalence of Higher-Order MMRs}
\label{sec: 4.3}

At the end of our simulations, we produced a total of $22,444$ pairs of planets in first-order MMRs, $1124$ in second-order MMRs, and $151$ in third-order MMRs (Fig. \ref{fig: 3}). Second- and third-order MMRs correspond to $4.74 \pm 0.13\%$ and $0.637 \pm 0.052\%$ of the produced MMRs. If, instead, we count the number of planetary systems that contain at least one higher-order MMR, we found that $720/5494$ or $13.03\pm0.45\%$ contain at least one second-order resonance, $98/5494$ or $1.77\pm0.18\%$ contain at least one third-order resonance. 

Focusing on systems where some planets are not incorporated in a resonant chain (labeled `Partial Resonant Chains' in Section \ref{sec: 2.3}), second- and third-order resonances occur at higher rates: $20.0\pm1.3\%$ and $2.6\pm0.5\%$ respectively. This difference is because the migration rate is generally slower in Partial Resonant Chains as some planets have not reached the inner disk yet. As demonstrated in Section \ref{sec: 4.4}, higher-order MMRs prefer slower migration.

As we cautiously noted above, these fractions of higher-order resonance critically depend on the prior range of disk surface density we assumed ($10-10,000\text{ g~cm}^{-2}$, see Section \ref{sec: 2.1}). Observational constraints on the surface densities of the innermost 1AU of protoplanetary disk are lacking \citep[e.g.][]{Andrews_review}. The more robust results from our simulations are the relative proportion of individual resonances e.g. the fraction of planets in 5:3 v.s. 7:5 MMR.

A complete list of the two-body resonances and their relative frequencies are shown in Tab. \ref{tbl: 4} in the Appendix. These fractions depend on the distribution of initial period ratios, which is why we selected a wide period ratio space (effectively 1.1-10) that mimics the non-resonant Kepler sample. In Fig. \ref{fig: 9}, we compare the frequencies of individual MMRs in our simulations with the observed sample reported in \citet{dai2024prevalence}. We find good agreement between the simulations and observations. Notably, within each order of MMR, the resonances with smaller period ratios (defined $P_{out}/P_{in}$) are increasingly rare, both in simulations and observations. This is because the planets have to avoid being captured into all preceding resonances before reaching the deeper resonances \citep{kajtazi2023mean}. For instance, both example systems in Section \ref{sec: 3} capture into the resonance with the next largest period ratio after breaking from a first-order MMR. 5:3 is the most populated resonance in both simulations and observations as this is the first strong and stable second-order resonance. \citet{Steffen2015} also noted 5:3 is the most prominent second-order resonance among the Kepler sample \citep[see also][]{Bailey}. 

The initial period ratios in our simulations influence which resonance planets encounter first. Planet pairs that start with larger period ratios tend to encounter the 3:1 or 2:1 MMR initially while those with tight period ratios may encounter resonances with smaller commensuralities like 6:5 or 7:6 first. However, the ability of those MMRs to capture and retain planet pairs also matters. At least half of deep MMRs are formed by planets that started with initial period ratios larger than 2 and avoided being captured into all preceding MMRs. The final outcome of our simulations is thus the combined effect of both initial conditions and resonant dynamics.

Crucially, as described in Section \ref{sec: 4.5}, most planet pairs undergo significant differential migration (some pairs initially go through a phase of divergent migration before they reach the inner disk edge and migrate convergently) that dramatically changes their initial period ratio. In other words, planet pairs that ended up in a particular MMR might not have started with a period ratio near that MMR. Therefore, the final distribution of various MMRs in our simulations are not solely determined by the initial period ratios we set. Instead, the distribution is driven primarily by the ability of individual MMRs to capture the sub-Neptune-mass planets and their relative migration rate assumed in our simulations.

\begin{figure*}
\begin{center}
\includegraphics[width = 2.\columnwidth]{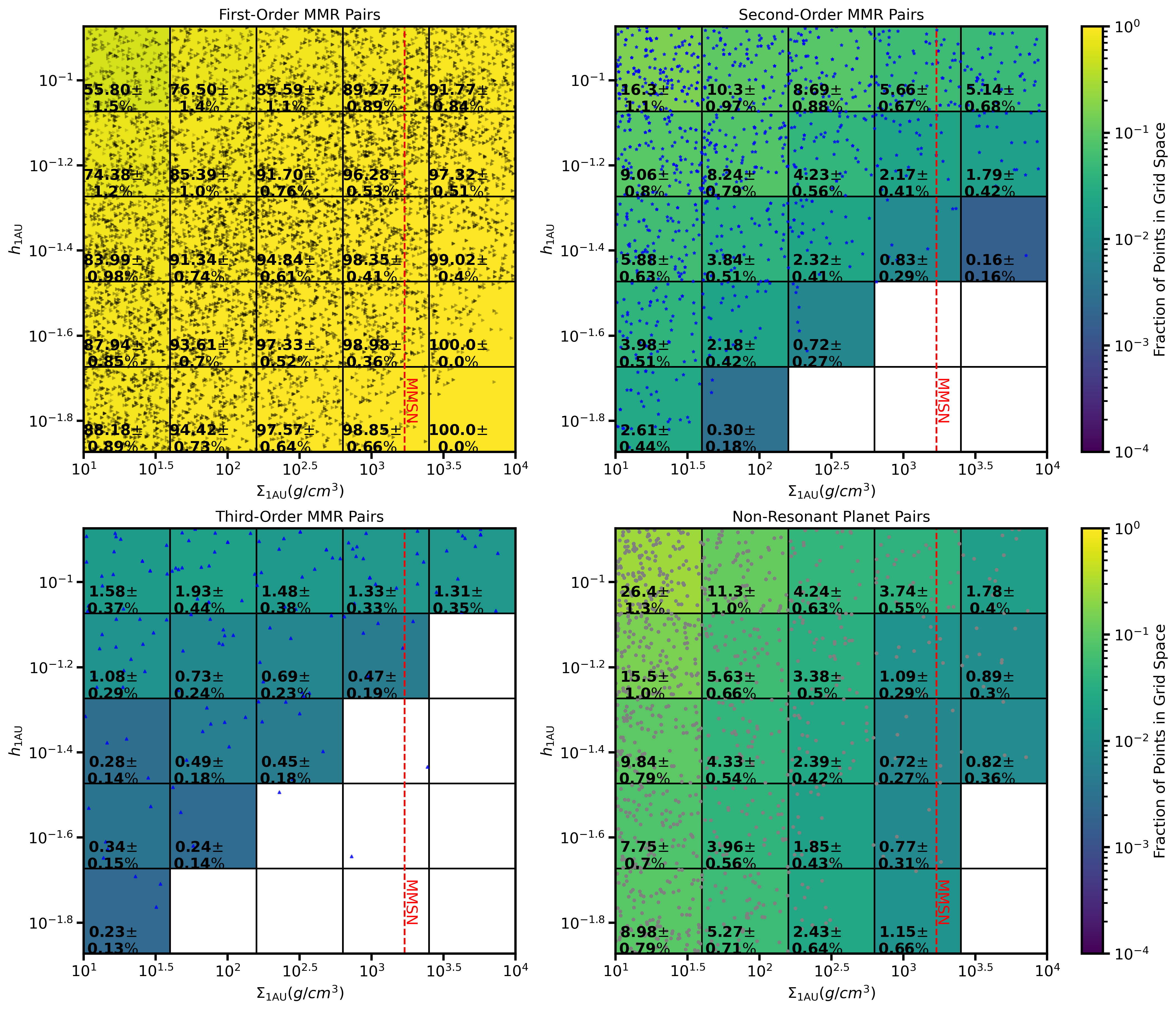}
\caption{The relative outcome of Type-I disk migration as a function of disk surface density ($\Sigma_{\rm 1AU}$) and disk aspect ratio ($h_{\rm 1AU}$). The four panels show planets that captured into first-, second-, and third-order MMR as well as non-resonant planets. For reference, a dotted red line marks the surface density of the Minimum Mass Solar Nebula \citep{Hayashi}. Higher-order resonances prefer low $\Sigma_{\rm 1AU}$ and high $h$, both of which slow down migration (Equation \ref{eq: 5}). Low $\Sigma_{\rm 1AU}$ may correspond to transitional disks or truncated disks; the formation of Kepler-like planets in such disks has been proposed previously \citep{Lee2016transitional,Dupuy444}.}
\label{fig: 11}
\end{center}
\end{figure*}

 Curiously, the $2$:$1$ and $3$:$1$ MMRs are more common in our simulations than in the observed sample. 2:1 even dominates over 3:2, which is the most prevalent observed first-order resonance. One likely source of this discrepancy is the relevant initial conditions. If real planetary systems start closer to each other than what we assumed in our initial conditions, more pairs would never encounter 2:1 and 3:1, naturally reducing their occurrence rates. This would be consistent with other observational evidence that low-mass planets may initially form in compact chains \citep{xu2024earths}. In other words, our simulations probably do not overestimate the probability for a system to stay resonant after encountering a 2:1 or 3:1 resonance, but we may overestimate the probability for a system to encounter 2:1 and 3:1 in the first place.

This explanation leaves one question: how can our initial period ratio condition be too wide when it was set according to the present-day observations, which, presumably, include planets that have undergone some convergent migration and are thus more compact than the true initial conditions? There are two factors that can contribute to this. First, middle planets may be missed in transit observations since planets in the same system have small but finite mutual inclinations \citep[e.g.][]{Zhu}. In other words, an adjacent planet pair with a large period ratio may actually have an intermediate planet, thus decreasing the reported period ratios. Second, a middle planet can be removed (colliding and merging with other planets or escaping the system) through long-term dynamical instability, which may take place during the Gyr evolution after disk dispersal \citep{Izidoro}. In these cases, the final separations could be larger than the initial conditions. For instance, \citet{thomas2025biases} found that removing planets from simulated and observed exoplanetary systems tend to increase the irregularity of planet spacings.

\subsection{Higher-Order MMRs Prefer Slower Migration} 
\label{sec: 4.4}

The probability of resonance capture strongly increases under adiabatic encounters i.e. the migration timescale should be slower than the resonant interaction timescale \citep[e.g.][]{henrard1982capture, henrard1983second, Batygin_resonance}. Theoretically, we anticipate that the formation of weaker, higher-order MMRs, which have narrower libration widths and slower libration timescales, demands slower migration than first-order. The differential migration rate between a pair of planets is set by the mass ratios between the planets (heavier planets tend to migrate faster) and the local disk properties (Equation \ref{eq: 3}).

In Fig. \ref{fig: 10}, we show the cumulative distribution functions of the mass ratios between neighboring planets and the migration rates for planets that ended up in MMRs of different orders. The mass ratio of neighboring planets $m_{\rm out}/m_{\rm in}$ are consistent with being around unity $log(m_{\rm out}/m_{\rm in})=0.0\pm0.4$ for pairs that ended up in first-order resonances, while the distribution for higher-order resonance shows a subtle preference smaller than unity $log(m_{\rm out}/m_{\rm in})=-0.1\pm0.4$. Indeed, a smaller outer planet leads to slower differential migration between the planets and thus favors capture into resonance. However, given how subtle the difference is, we do not expect any observable difference in the mass ratios for planets in higher-order resonances from those in first-order resonances. \citet{Goyal} reported that intra-system uniformity in planetary mass appears to be stronger in observed resonant systems i.e. neighboring planets are similar in size.

\begin{figure*}
\begin{center}
\includegraphics[width = 2.\columnwidth]{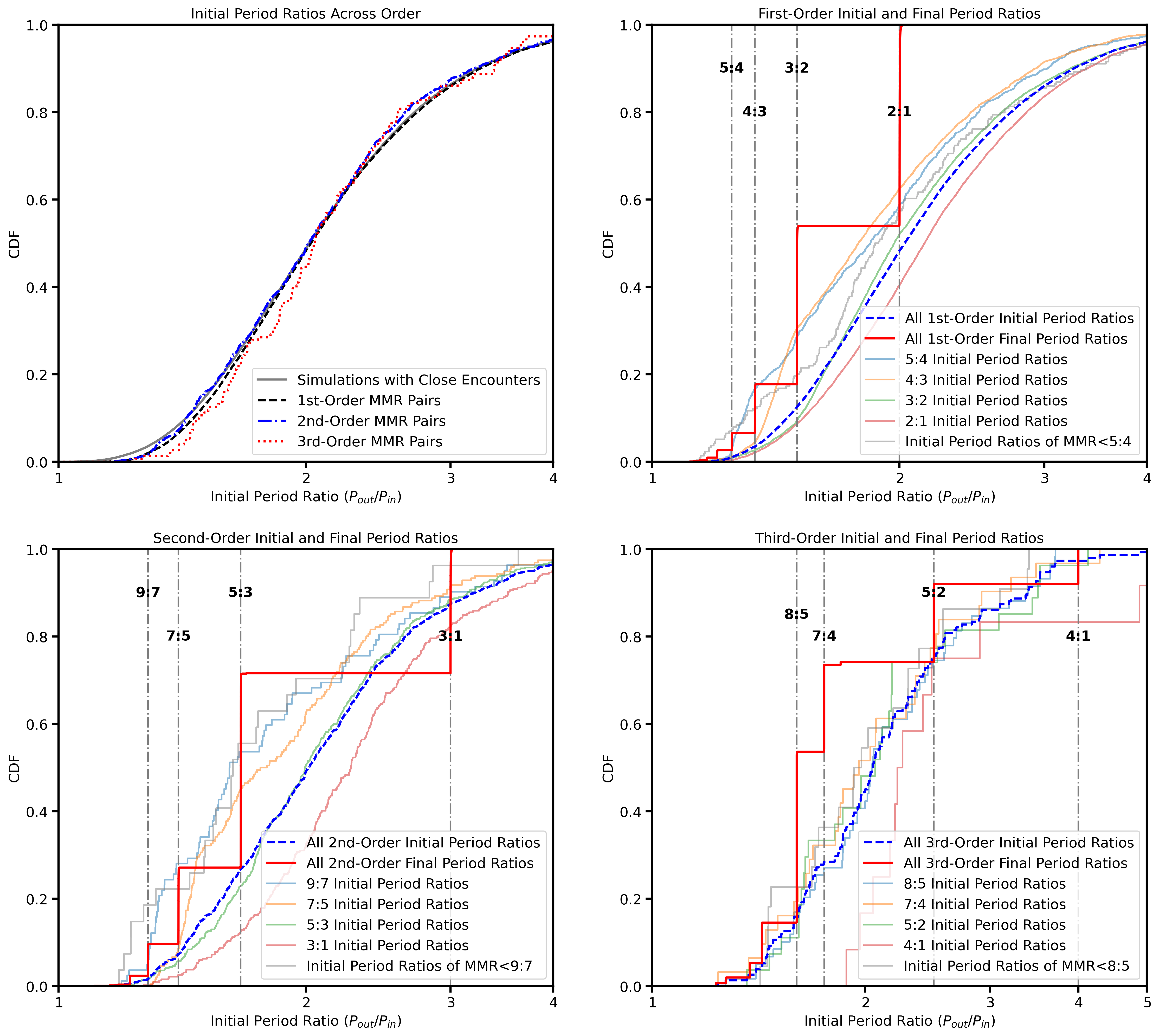}
\caption{Top left: CDF of the initial period ratios of the planet pairs that captured into first-, second-, and third-order MMR. The remaining three panels show the initial period ratios and final period ratios for each order. To provide further context, we display CDFs of the initial period ratios for the four most populated MMRs of each order as well as a CDF of the initial period ratios for all remaining MMRs of that order.} Notice that a planet pair that end up in a higher-order MMR need not begin with a commensurate period ratio.
\label{fig: 12}
\end{center}
\end{figure*}

\citet{Xu0217} also proposed that planets in second-order MMRs should have order unity mass ratios: $m_{\text{out}}/m_{\text{in}} \sim 1$. This result follows because, on the one hand, convergent migration requires a massive outer planet ($m_{\text{out}}/m_{\text{in}} > 1$), but, on the other, the stability of a captured resonance demands $m_{\text{out}}/m_{\text{in}} < 1$ to avoid overstable libration. Our simulations are broadly consistent with this result, and the stability criterion may have contributed to the preference for lower $m_{\text{out}}/m_{\text{in}}$, but our mass ratio distribution is wider: it spans almost half a dex ($log(m_{\rm out}/m_{\rm in})=-0.1\pm0.4$). We argue that this is because the disk inner edge was crucial for converting divergent encounters into convergent ones by stopping the migration of the inner planets, making it possible for a less massive outer planet to catch up. Moreover, in a resonant chain, neighboring resonances may also help stabilize a higher-order MMR e.g. through three-body Laplace-like resonance \citep{Agol2021}. Indeed, $79.3\%$ of higher-order two-body MMRs participate in at least one three-body resonance. \citet{Xu0217} only considered isolated pairs of planets.

\begin{figure}
    \begin{center}
    \includegraphics[width = \linewidth]{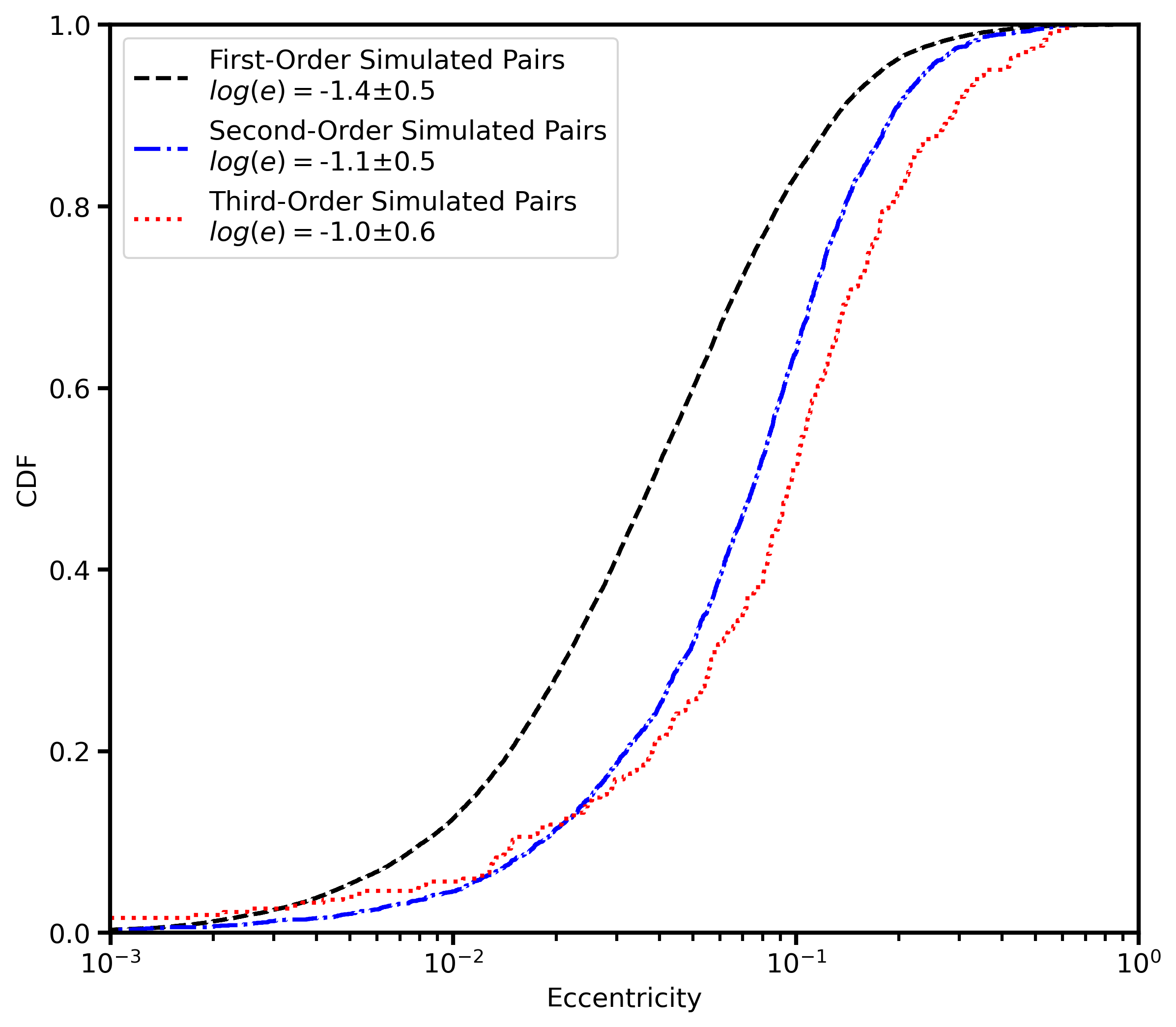}
    \caption{CDFs of eccentricities for planets that captured into first-, second-, and third-order MMRs. Higher-order MMRs have larger $e$ than first-order resonances (KS tests confirmed the statistical significance).}
    \label{fig: 13}
    \end{center}
\end{figure}

Slower migration favors the capture into higher-order resonances. The average migration $\tau_a$ is larger for higher-order resonance than first-order: $\tau_a=10^{3.0\pm1.0}$ kyr v.s. $\tau_a=10^{2.4\pm1.0}$ kyr (see Tab. \ref{tbl: 3} and Fig. \ref{fig: 10}). Although the two distributions overlap substantially, our simulated sample of 6000 systems provide enough statistical power to differentiate them in a Kolmogorov-Smirnov (KS) test \citep{berger2014kolmogorov} ($5\sigma$). Translated to $\Sigma$ and $h$, higher-order resonances prefer lower disk surface densities and larger disk aspect ratio: $\Sigma_{\rm 1AU} = 10^{1.8\pm0.8}$g~cm$^{-2}$ v.s.$\Sigma_{\rm 1AU} = 10^{2.3\pm0.8}$g~cm$^{-2}$; $h = 10^{-1.0\pm0.2}$ v.s. $h = 10^{-1.3\pm0.3}$. In Fig. \ref{fig: 11}, we display the 2-D parameter space of $\Sigma$-$h$ and show where higher-order resonances tend to emerge. The MMSN \citep{Hayashi} is labeled for comparison.

\subsection{Higher-Order MMRs do not Require Commensurate Initial Period Ratios} 
\label{sec: 4.5}

One might naively expect that the planets that end up in higher-order MMR were initialized with a period ratio commensurate with the final resonance, avoiding intervening stronger first-order MMRs. After a short migration, these planets could have captured into the nearby higher-order MMR. In previous simulations of resonant chains, this approach was adopted to capture planets into a series of pre-determined resonances \citep[e.g.,][]{Huang_Ormel, Pichierri2018, Tamayo2017, lammers2024six}. 

In our simulations, we found that higher-order MMRs do not require initial period ratios that are commensurate with the final resonance. Specifically, only $43$ out of the second-order $1124$ pairs started with a period ratio within 2\% of the final resonance. Similarly, only $6$ out of the $151$ third-order pairs started with near-commensurate period ratios. In Fig. \ref{fig: 12}, we show the cumulative distributions of the initial period ratios for all planet pairs that end up in first-, second-, and third-order MMRs separately. We found the different orders had statistically indistinguishable initial period ratios. The p-values from a KS test \citep{berger2014kolmogorov} between the first and second-order was $0.2$, and between first- and third-order were $0.4$. Moreover, there are no discernible peaks in the initial period distribution near the final resonances (Fig. \ref{fig: 12}). When the initial period ratios for pairs that end up in a specific MMR are examined individually, pileups can be present near the position of the MMR, especially for deeper resonances. However, in all of these cases, a plethora of pairs still begin external to the $2{:}1$ resonance.

The explanation is simple: most higher-order MMRs have to undergo substantial migration. Often, as seen in the example systems described in Section \ref{sec: 3}, planets experience significant divergent migration away from their initial period ratios. After the inner planet reaches the disk inner edge, they begin convergent migration. The pair likely briefly resided in a first-order MMR before breaking away and capturing into a nearby higher-order resonance as discussed in the case studies of Section \ref{sec: 3}. A quick examination of $\sim 100$ systems suggests that around half of higher-order MMRs form through this pathway. In this scenario, the initial period ratios are a less significant driver of the final MMR than other initial conditions, like the disk surface density and the planet-planet mass ratios.

\subsection{Higher-Order MMRs Have Higher Eccentricity} 
\label{sec: 4.6}

In our simulations, the planets in higher-order MMRs have larger equilibrium eccentricities than their first-order counterparts. This result is shown in Fig. \ref{fig: 13}, where KS tests indicate that the $e$ distributions of higher-order MMRs are statistically different from that of first-order at $>5\sigma$. The eccentricities are log$(e)=-1.4\pm0.5$ for first-order MMR, log$(e)=-1.1\pm0.5$ for second-order, and log$(e)=-1.0\pm0.6$ for third-order. Again, this is because slower eccentricity damping imparts a non-zero eccentricity before planets encounter higher-order MMR and favors capture (see Section \ref{sec: 4.2}).
 
The high eccentricities for higher-order MMRs may lead to resonance overlap and orbital instability after disk dispersal \citep[e.g.][]{lammers2024instability, petit2020resonance, tamayo2021criterion, Hadden2018, deck2013first}. Higher-order MMR may be the weakest link of a resonant chain and contribute to the breaking of chains \citep{dai2023toi}. We defer a dynamical stability analysis of our simulated higher-order MMR to a future work.

\subsection{Innermost Pairs More Likely Form Higher-Order MMRs}
\label{sec: 4.7}

We found that higher-order MMRs tend to form on the innermost planet pair of a resonant chain (see example systems in Fig. \ref{fig: 4}). Given the adopted multiplicity of planets (3-7) in our simulations, the innermost pairs represent $25.9\%$ of all neighboring pairs. The formation of strong first-order resonance seem to be agnostic about the relative location of the planet pair in a resonant chain: the innermost pair account for $5786/22444=25.8\%$ of first-order MMR.  The fraction of innermost pairs engaged in a higher-order MMR is significantly higher, $419/1124=37.3\%$ and $63/151=41.7\%$ for second- and third-order MMR respectively.


Planets in the innermost pair are more likely to break from the first MMR because this pair of planets are pushed from both sides in our simulations. The innermost planet is typically at the disk inner edge, so the net migration is outward. The second to innermost planet migrates inwards. Moreover, there is often a whole chain of planets locked in resonance whose net migration are all inwards. As the innermost pair is squeezed, a previously established first-order resonance can break and subsequent migration can capture that pair into a nearby higher-order MMR. This pattern is essentially what happened in the case studies in Section \ref{sec: 3}. That said, longer-period planets can still form higher-order MMR. This is both seen in our simulations and in observations, e.g. TOI-1136 ef \citep{dai2023toi}.

\begin{deluxetable*}{ccc}
\tablecaption{Factors that Promote Higher-order MMRs}\label{tbl: 1}
\tablehead{
\colhead{Factor} & \colhead{Higher-order MMR} & \colhead{First-order MMR}}
\startdata
Larger $log(\tau_a/kyr)$ & $3.0\pm1.0$ & $2.4\pm1.0$ \\
Smaller $log(m_{out}/m_{in})$ & $-0.1\pm0.5$ & $0.0\pm0.4$ \\
Smaller $log(\Sigma/g~cm^{-2})$ & $1.8\pm0.8$ & $2.3\pm0.8$ \\
Larger $log(h)$ & $-1.1\pm0.2$ & $-1.3\pm0.3$ \\
Smaller $log(K)$ & $1.5\pm0.5$ & $2.0\pm0.5$ \\
\enddata
\end{deluxetable*}

\section{Conclusion} 
\label{sec: 5}

In this paper, we investigated the formation of second- and third-order MMRs during Type-I migration with an inner disk edge. We ran more than $6000$ simulations using the {\tt type\_I\_migration} \citep{kajtazi2023mean} scheme in {\tt REBOUNDx} \citep{Tamayo_x,Rein}. Our simulations aimed to reproduce the observed stellar mass, planet radius/mass, multiplicity, and intra-system uniformity of Kepler-like planets \citep[e.g.][]{Fabrycky2014,Weiss_peas,Millholland,Wang_uniform,Zhu_Dong}. We included a wide set of protoplanetary disk surface densities $10-10^4$ g~cm$^{-2}$ at 1AU and aspect ratios $H/R=0.1-0.01$. This range may encompass transitional or truncated disks \citep{Lee2016transitional,Dupuy444} i.e. during the last stage of planet formation when the resonant chains are assembled. Our findings are as follows:

\begin{enumerate} 
    \item Among $>6000$ simulated systems, $\sim 5\%$ and $\sim 0.5\%$ of resonant planet pairs were captured into second- and third-order MMRs (in a state of libration); $\sim 13\%$ and $\sim 2\%$ of systems contain at least one second-order or third-order MMR.
    \item Even though the above fractions depended on the assumed disk properties, the fraction of individual resonances (e.g. 5:3 v.s. 7:5) in our simulations reproduced that of the observed sample \citep{dai2024prevalence} very well. MMRs with tight period ratio spacings are increasingly rare (Tab. \ref{tbl: 4}) as planets have to avoid capture into all external MMRs. $2$:$1$ and $3$:$1$ resonances are significantly more common in our simulations than in the observed sample, likely because of observational biases.
    \item As predicted by theory \citep{Xu0217}, higher-order MMRs more likely emerge in lower-density disks  ($\Sigma_{\rm 1AU} = 10^{1.8\pm0.8}$g~cm$^{-2}$ v.s. $\Sigma_{\rm 1AU} = 10^{2.3\pm0.8}$g~cm$^{-2}$ for first-order). The distinction in disk density is small enough that we expect to find higher-order MMRs in the same observational sample as first-order resonances.
    \item A pair of planets can capture into a higher-order MMR even (and perhaps especially) when there is an intervening first-order MMR. Specifically, only $43$ out of the $1124$ second-order pairs and $6$ of $151$ third-order pairs started with a period ratio within 2\% of the final resonance. Planets typically enter a ``stronger'' first-order MMR before breaking away and gently capturing into a nearby higher-order MMR. The initial period ratio does not solely define the order of the MMR.
    \item Higher-order MMR do not have to form as part of a pre-existing Laplace-like three-body resonance. The majority of our higher-order MMRs form through two-body resonances.
    \item Instead, we suggest that small but non-zero pre-capture eccentricities caused by a prior first-order resonance facilitate the capture into higher-order MMR.
\end{enumerate}

Based on our simulations, we also make some predictions about higher-order MMRs in observed systems:
\begin{enumerate}
    \item The formation of higher-order MMRs prefers disks that cause slow migration. Many of the simulated systems with higher-order MMRs have longer-period planets that have yet to reach the inner disk whereas rapid disk migration mostly gives rise to complete first-order resonant chains where all planets have finished migration. We predict that higher-order MMR planets are more likely to occur in systems with longer-period non-resonant planets.
    \item The absolute frequency of higher-order MMRs depends on disk properties, but higher-order MMRs are rare. They should tend to appear as an isolated pair in an otherwise first-order resonant chain.
    \item The inner pairs of a resonant chain are more likely engaged in higher-order MMR. This is because outward migration on the innermost planet and inward migration of all longer-period planets tend to squeeze this pair and break it. Breaking from resonance gives the planets another chance to capture into nearby higher-order resonance.
    \item Slower eccentricity damping help maintain a non-zero pre-capture eccentricities which in turn facilitate the capture into higher-order MMR. In our simulations, higher-order MMRs tend to have higher equilibrium eccentricities: log$(e)=-1.4\pm0.5$ for first-order MMR, log$(e)=-1.1\pm0.5$ for second-order, and log$(e)=-1.0\pm0.6$ for third-order resonance. Such high eccentricities for higher-order resonances may lead to resonance overlap and orbital instability after the disk dissipates \citep[e.g.][]{lammers2024instability, petit2020resonance, tamayo2021criterion, Hadden2018, deck2013first}. We predict young planets in higher-order MMR may have $e$ as high as 0.1, higher than that of mature planets \citep[e.g. $\approx0.05$]{Hadden2017} and their first-order counterparts.
\end{enumerate}

We acknowledge that effects such as disk turbulence \citep{Adams2008,Goldberg2023,Wu_Chen_Lin2024, chen2025capture}, disk evolution \citep{pichierri2024formation, Hansen2024, Huang_Ormel, liu2017dynamical}, the collisional growth/gas accretion of planets \citep{Izidoro}, and density wave interaction \citep{yang2024mean} have not been accounted for in our simulations. Post-formation dynamical evolution of these systems is the obvious next step. Higher-order MMRs may play an important role in the disruption of resonant chains \citep{Pichierri2020, Goldberg2022, Izidoro,Goldberg_stability,Li2024}. In particular, the enhanced eccentricities of higher-order MMR planets may quickly go unstable due to resonance overlap \citep{Hadden2018}. Observationally, \cite{hamer2024kepler} already found evidence that second-order MMRs are short-lived. We defer a direct N-body demonstration of this effect to a future study. 

\software{{\tt REBOUND} \citep{Rein}, {\tt REBOUNDx} \citep{Tamayo_x}, {\tt celmech} \citep{hadden2022celmech}, {\tt forecaster} \citep{Chen2017}, {\tt pandas} \citep{reback2020pandas, mckinney-proc-scipy-2010}, {\tt numpy} \citep{harris2020array}, {\tt scipy} \citep{2020SciPy-NMeth}, {\tt astropy} \citep{astropy:2022}, {\tt Matplotlib} \citep{Hunter:2007}, {\tt Seaborn} \citep{Waskom2021}, {\tt label--lines} \citep{https://doi.org/10.5281/zenodo.7428070}}

\acknowledgements
We thank Caleb Lammers, Max Goldberg, Daniel Tamayo, Nick Choksi, Renu Malhotra, Zhecheng Hu, Tian Yi, Shuo Huang, Konstantin Batygin, Mutian Wang, Eric Agol, Rixin Li, Fred Adams, and Daniel Fabrycky for useful conversations and suggestions. To provide observational radii and masses, this research has made use of the NASA Exoplanet Archive, which is operated by the California Institute of Technology, under contract with the National Aeronautics and Space Administration under the Exoplanet Exploration Program. Most of the dynamical simulations were conducted using computational resources and services at the Center for Computation and Visualization, Brown University. FMK thanks Gregory Tucker and the Undergraduate Teaching and Research Award at Brown University for their support of his undergraduate research.

\bibliography{main}
\newpage

\appendix 

We provide the parameters for the second-order and third-order example systems described in Section \ref{sec: 3} alongside the third-order evolution plot. Additionally, we list the abundances of each resonance observed in our simulations, as detailed in Section \ref{sec: 2.4}.

\begin{centering}
\begin{deluxetable*}{ccccccc}[ht]
\tablecaption{System Parameters}\label{tbl: 2}
\tablehead{
\colhead{\textcolor{red}{Planet $0$}} & \colhead{\textcolor{red}{Planet $1$}} & \colhead{Planet $2$} & \colhead{Planet $3$} & \colhead{Planet $4$} & \colhead{Planet $5$} & \colhead{Planet $6$}
}
\startdata
\textcolor{red}{$m_0$} & \textcolor{red}{$m_1$} & $m_2$ & $m_3$ & $m_4$ & $m_5$ & $m_6$ \\
\textcolor{red}{$7.95M_\oplus$} & \textcolor{red}{$4.82M_\oplus$} & $8.64M_\oplus$ & $8.55M_\oplus$ & $8.59M_\oplus$ & $4.42M_\oplus$ & $8.67M_\oplus$ \\
\hline
\textcolor{red}{$e_0$} & \textcolor{red}{$e_1$} & $e_2$ & $e_3$ & $e_4$ & $e_5$ & $e_6$ \\
\textcolor{red}{$0.0507$} & \textcolor{red}{$0.175$} & $0.0492$ & $0.0426$ & $0.0442$ & $0.0251$ & $0.0121$ \\
\hline
\textcolor{red}{$a_{0i}$} & \textcolor{red}{$a_{1i}$} & $a_{2i}$ & $a_{3i}$ & $a_{4i}$ & $a_{5i}$ & $a_{6i}$ \\
\textcolor{red}{$0.1AU$} & \textcolor{red}{$0.173AU$} & $0.29AU$ & $0.531AU$ & $0.644AU$ & $1.08AU$ & $1.37AU$ \\
\hline
\textcolor{red}{$a_{0f}$} & \textcolor{red}{$a_{1f}$} & $a_{2f}$ & $a_{3f}$ & $a_{4f}$ & $a_{5f}$ & $a_{6f}$ \\
\textcolor{red}{$0.0501AU$} & \textcolor{red}{$0.0705AU$} & $0.112AU$ & $0.178AU$ & $0.233AU$ & $0.37AU$ & $0.448AU$ \\
\hline
\hline
\textcolor{red}{Planet Pair $01$} & Planet Pair $12$ & Planet Pair $23$ & Planet Pair $34$ & Planet Pair $45$ & Planet Pair $56$ & \\
\hline 
\textcolor{red}{$m_1/m_0$} & $m_2/m_1$ & $m_3/m_2$ & $m_4/m_3$ & $m_5/m_4$ & $m_6/m_5$ & \\
\textcolor{red}{$0.607$} & $1.79$ & $0.989$ & $1.0$ & $0.515$ & $1.96$ & \\
\hline
\textcolor{red}{$p_{01}/q_{01}$} & $p_{12}/q_{12}$ & $p_{23}/q_{23}$ & $p_{34}/q_{34}$ & $p_{45}/q_{45}$ & $p_{56}/q_{56}$ & \\
\textcolor{red}{$5/3$} & $2$ & $2$ & $3/2$ & $2$ & $4/3$ & \\
\hline
\textcolor{red}{$(P_1/P_0)_f$} & $(P_2/P_1)_f$ & $(P_3/P_2)_f$ & $(P_4/P_3)_f$ & $(P_5/P_4)_f$ & $(P_6/P_5)_f$ & \\
\textcolor{red}{$1.6668$} & $2.0004$ & $2.0007$ & $1.5006$ & $2.0021$ & $1.3344$ & \\
\hline
\textcolor{red}{$\Delta_{01}$} & $\Delta_{12}$ & $\Delta_{23}$ & $\Delta_{34}$ & $\Delta_{45}$ & $\Delta_{56}$ & \\
\textcolor{red}{$8.0737\cdot10^{-5}$} & $0.00017715$ & $0.00037187$ & $0.00039041$ & $0.0010665$ & $0.00078198$ & \\
\hline
\textcolor{red}{$\phi_{01}$} & $\phi_{12}$ & $\phi_{23}$ & $\phi_{34}$ & $\phi_{45}$ & $\phi_{56}$ & \\
\textcolor{red}{$180.0^\circ$} & $167.0^\circ$ & $201.0^\circ$ & $180.0^\circ$ & $189.0^\circ$ & $179.0^\circ$ & \\
\hline
\textcolor{red}{$A_{01}$} & $A_{12}$ & $A_{23}$ & $A_{34}$ & $A_{45}$ & $A_{56}$ & \\
\textcolor{red}{$0.184^\circ$} & $0.0631^\circ$ & $0.0656^\circ$ & $0.194^\circ$ & $0.136^\circ$ & $1.23^\circ$ & \\
\hline
\hline
\textcolor{red}{Planet Triplet $012$} & Planet Triplet $123$ & Planet Triplet $234$ & Planet Triplet $345$ & Planet Triplet $456$ & & \\
\hline
\textcolor{red}{$\phi_{012}$} & $\phi_{123}$ & $\phi_{234}$ & $\phi_{345}$ & $\phi_{456}$ & & \\
\textcolor{red}{$196.0^\circ$} & $39.7^\circ$ & $237.0^\circ$ & $180.0^\circ$ & $242.0^\circ$ & & \\
\textcolor{red}{$A_{012}$} & $A_{123}$ & $A_{234}$ & $A_{345}$ & $A_{456}$ & & \\
\textcolor{red}{$0.272^\circ$} & $0.0484^\circ$ & $0.0873^\circ$ & $0.127^\circ$ & $0.326^\circ$ & & \\
\hline
\hline
$\Sigma_{\rm 1AU}$ & $h$ & $\tau_a$ & Integration Time & Stellar Mass & & \\
$16.6g/cm^2$ & $0.0395$ & $1350kyr$ & $4050kyr$ & $0.89M_\odot$ & & \\
\enddata
\tablecomments{Orbital, migration, and MMR parameters for each planet in the example second-order system. Parameters of the planets involved in the second-order resonance are highlighted in red.}
\end{deluxetable*}
\end{centering}

\begin{figure*}[h]
\begin{center}
\includegraphics[width = 0.8\linewidth]{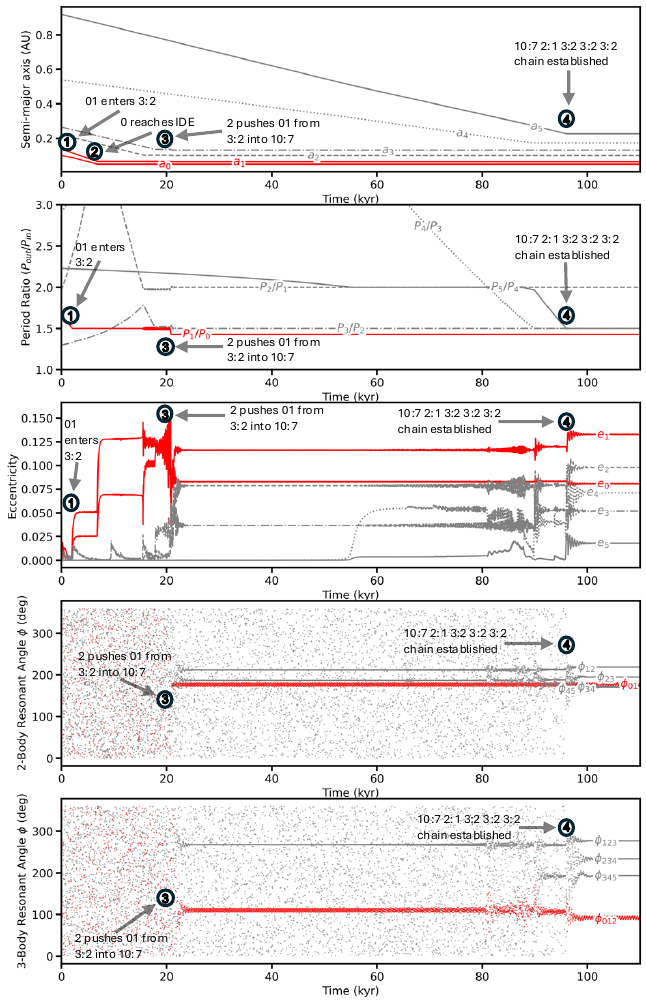}
\caption{Migration history of a system that ended up with a 10:7 third-order resonance. The innermost two planets entered a first-order $3$:$2$ resonance at milestone (1). At (2), the pair of planets stopped at the inner disk edge. The innermost planet pair broke away from the $3$:$2$ resonance at (3) and quickly settled into the third-order $10$:$7$ resonance. The system eventually became a complete resonant chain at (4).}
\label{fig: 14}
\end{center}
\end{figure*}

\begin{centering}
\begin{deluxetable*}{cccccc}[h]
\tablecaption{System Parameters}\label{tbl: 3}
\tablehead{
\colhead{\textcolor{red}{Planet $0$}} & \colhead{\textcolor{red}{Planet $1$}} & \colhead{Planet $2$} & \colhead{Planet $3$} & \colhead{Planet $4$} & \colhead{Planet $5$}
}
\startdata
\textcolor{red}{$m_0$} & \textcolor{red}{$m_1$} & $m_2$ & $m_3$ & $m_4$ & $m_5$ \\
\textcolor{red}{$3.07M_\oplus$} & \textcolor{red}{$6.73M_\oplus$} & $4.03M_\oplus$ & $3.86M_\oplus$ & $2.09M_\oplus$ & $3.93M_\oplus$ \\
\hline
\textcolor{red}{$e_0$} & \textcolor{red}{$e_1$} & $e_2$ & $e_3$ & $e_4$ & $e_5$ \\
\textcolor{red}{$0.0808$} & \textcolor{red}{$0.133$} & $0.0979$ & $0.052$ & $0.0709$ & $0.0181$ \\
\hline
\textcolor{red}{$a_{0i}$} & \textcolor{red}{$a_{1i}$} & $a_{2i}$ & $a_{3i}$ & $a_{4i}$ & $a_{5i}$ \\
\textcolor{red}{$0.1AU$} & \textcolor{red}{$0.141AU$} & $0.222AU$ & $0.264AU$ & $0.538AU$ & $0.919AU$ \\
\hline
\textcolor{red}{$a_{0f}$} & \textcolor{red}{$a_{1f}$} & $a_{2f}$ & $a_{3f}$ & $a_{4f}$ & $a_{5f}$ \\
\textcolor{red}{$0.0499AU$} & \textcolor{red}{$0.0633AU$} & $0.1AU$ & $0.132AU$ & $0.173AU$ & $0.226AU$ \\
\hline
\hline
\textcolor{red}{Planet Pair $01$} & Planet Pair $12$ & Planet Pair $23$ & Planet Pair $34$ & Planet Pair $45$ & \\
\hline 
\textcolor{red}{$m_1/m_0$} & $m_2/m_1$ & $m_3/m_2$ & $m_4/m_3$ & $m_5/m_4$ & \\
\textcolor{red}{$2.19$} & $0.599$ & $0.959$ & $0.541$ & $1.88$ & \\
\hline
\textcolor{red}{$p_{01}/q_{01}$} & $p_{12}/q_{12}$ & $p_{23}/q_{23}$ & $p_{34}/q_{34}$ & $p_{45}/q_{45}$ & \\
\textcolor{red}{$10/7$} & $2$ & $3/2$ & $3/2$ & $3/2$ & \\
\hline
\textcolor{red}{$(P_1/P_0)_f$} & $(P_2/P_1)_f$ & $(P_3/P_2)_f$ & $(P_4/P_3)_f$ & $(P_5/P_4)_f$ & \\
\textcolor{red}{$1.4286$} & $2.0001$ & $1.5001$ & $1.5002$ & $1.5002$ & \\
\hline
\textcolor{red}{$\Delta_{01}$} & $\Delta_{12}$ & $\Delta_{23}$ & $\Delta_{34}$ & $\Delta_{45}$ & \\
\textcolor{red}{$4.886\cdot10^{-5}$} & $6.4261\cdot10^{-5}$ & $8.2549\cdot10^{-5}$ & $0.00013246$ & $0.00016031$ & \\
\hline
\textcolor{red}{$\phi_{01}$} & $\phi_{12}$ & $\phi_{23}$ & $\phi_{34}$ & $\phi_{45}$ & \\
\textcolor{red}{$177.0^\circ$} & $219.0^\circ$ & $195.0^\circ$ & $172.0^\circ$ & $170.0^\circ$ & \\
\hline
\textcolor{red}{$A_{01}$} & $A_{12}$ & $A_{23}$ & $A_{34}$ & $A_{45}$ & \\
\textcolor{red}{$3.89^\circ$} & $0.125^\circ$ & $0.0561^\circ$ & $0.0538^\circ$ & $0.0719^\circ$ & \\
\hline
\hline
\textcolor{red}{Planet Triplet $012$} & Planet Triplet $123$ & Planet Triplet $234$ & Planet Triplet $345$ & & \\
\hline
\textcolor{red}{$\phi_{012}$} & $\phi_{123}$ & $\phi_{234}$ & $\phi_{345}$ & & \\
\textcolor{red}{$92.3^\circ$} & $277.0^\circ$ & $234.0^\circ$ & $194.0^\circ$ & & \\
\textcolor{red}{$A_{012}$} & $A_{123}$ & $A_{234}$ & $A_{345}$ & & \\
\textcolor{red}{$4.19^\circ$} & $0.145^\circ$ & $0.05^\circ$ & $0.0569^\circ$ & & \\
\hline
\hline
$\Sigma_{\rm 1AU}$ & $h$ & $\tau_a$ & Integration Time & Stellar Mass & \\
$2080g/cm^2$ & $0.0991$ & $123kyr$ & $369kyr$ & $1.02M_\odot$ & \\
\enddata
\tablecomments{Orbital, migration, and MMR parameters for each planet in the example third-order system. Parameters of the planets involved in the third-order resonance are highlighted in red.}
\end{deluxetable*}
\end{centering}

\begin{centering}
\begin{deluxetable*}{lcc}
\caption{Frequencies of Individual MMR}\label{tbl: 4}
\tablehead{
    \colhead{Category} & \colhead{Count} & \colhead{Percentage Against Category}}
\startdata
First-Order Resonant Pairs & $22444$ & $94.6\pm0.1\%$\\
2:1 MMR & $10321$ & $46.0\pm0.003\%$\\
3:2 MMR & $8138$ & $36.3\pm0.003\%$\\
4:3 MMR & $2505$ & $11.2\pm0.002\%$\\
5:4 MMR & $885$ & $3.94\pm0.001\%$\\
6:5 MMR & $381$ & $1.7\pm0.0009\%$\\
7:6 MMR & $113$ & $0.503\pm0.0005\%$\\
8:7 MMR & $50$ & $0.223\pm0.0003\%$\\
9:8 MMR & $37$ & $0.165\pm0.0003\%$\\
10:9 MMR & $10$ & $0.0446\pm0.0001\%$\\
11:10 MMR & $4$ & $0.0178\pm9\cdot10^{-5}\%$\\
\hline
Second-Order Resonant Pairs & $1124$ & $4.74\pm0.1\%$\\
3:1 MMR & $319$ & $28.4\pm0.01\%$\\
5:3 MMR & $500$ & $44.5\pm0.01\%$\\
7:5 MMR & $196$ & $17.4\pm0.01\%$\\
9:7 MMR & $82$ & $7.3\pm0.008\%$\\
11:9 MMR & $23$ & $2.05\pm0.004\%$\\
13:11 MMR & $2$ & $0.178\pm0.001\%$\\
15:13 MMR & $1$ & $0.089\pm0.0009\%$\\
21:19 MMR & $1$ & $0.089\pm0.0009\%$\\
\hline
Third-Order Resonant Pairs & $151$ & $0.637\pm0.05\%$\\
4:1 MMR & $12$ & $7.95\pm0.02\%$\\
5:2 MMR & $27$ & $17.9\pm0.03\%$\\
7:4 MMR & $31$ & $20.5\pm0.03\%$\\
8:5 MMR & $59$ & $39.1\pm0.04\%$\\
10:7 MMR & $14$ & $9.27\pm0.02\%$\\
11:8 MMR & $5$ & $3.31\pm0.01\%$\\
14:11 MMR & $2$ & $1.32\pm0.009\%$\\
16:13 MMR & $1$ & $0.662\pm0.007\%$\\
\enddata
\tablecomments{The number of planet pairs that capture into each mean-motion resonance defined by $p$:$q$ observed in our simulations. Uncertainties are estimated from counting statistics.}
\end{deluxetable*}
\end{centering}

\end{document}